\documentstyle[]{article}

\catcode`\@=11
\long\def\@makefntext#1{
\protect\noindent \hbox to 3.2pt {\hskip-.9pt
$^{{\eightrm\@thefnmark}}$\hfil}#1\hfill}               

\def\@makefnmark{\hbox to 0pt{$^{\@thefnmark}$\hss}}    

\def\ps@myheadings{\let\@mkboth\@gobbletwo
\def\@oddhead{\hbox{}
\rightmark\hfil\eightrm\thepage}
\def\@oddfoot{}\def\@evenhead{\eightrm\thepage\hfil
\leftmark\hbox{}}\def\@evenfoot{}
\def\sectionmark##1{}\def\subsectionmark##1{}}

\oddsidemargin=\evensidemargin
\newcounter{sectionc}\newcounter{subsectionc}\newcounter{subsubsectionc}
\renewcommand{\section}[1] {\vspace{12pt}\addtocounter{sectionc}{1}
\setcounter{subsectionc}{0}\setcounter{subsubsectionc}{0}\noindent
        {\tenbf\thesectionc. #1}\par\vspace{5pt}}
\renewcommand{\subsection}[1] {\vspace{12pt}\addtocounter{subsectionc}{1}
        \setcounter{subsubsectionc}{0}\noindent
        {\bf\thesectionc.\thesubsectionc. {\kern1pt \bfit #1}}\par\vspace{5pt}}
\renewcommand{\subsubsection}[1] {\vspace{12pt}\addtocounter{subsubsectionc}{1}
        \noindent{\tenrm\thesectionc.\thesubsectionc.\thesubsubsectionc.
        {\kern1pt \tenit #1}}\par\vspace{5pt}}
\newcommand{\nonumsection}[1] {\vspace{12pt}\noindent{\tenbf #1}
        \par\vspace{5pt}}

\newcounter{appendixc}
\newcounter{subappendixc}[appendixc]
\newcounter{subsubappendixc}[subappendixc]
\renewcommand{\thesubappendixc}{\Alph{appendixc}.\arabic{subappendixc}}
\renewcommand{\thesubsubappendixc}
        {\Alph{appendixc}.\arabic{subappendixc}.\arabic{subsubappendixc}}

\renewcommand{\appendix}[1] {\vspace{12pt}
        \refstepcounter{appendixc}
        \setcounter{figure}{0}
        \setcounter{table}{0}
        \setcounter{lemma}{0}
        \setcounter{theorem}{0}
        \setcounter{corollary}{0}
        \setcounter{definition}{0}
        \setcounter{equation}{0}
        \renewcommand{\thefigure}{\Alph{appendixc}.\arabic{figure}}
        \renewcommand{\thetable}{\Alph{appendixc}.\arabic{table}}
        \renewcommand{\theappendixc}{\Alph{appendixc}}
        \renewcommand{\thelemma}{\Alph{appendixc}.\arabic{lemma}}
        \renewcommand{\thetheorem}{\Alph{appendixc}.\arabic{theorem}}
        \renewcommand{\thedefinition}{\Alph{appendixc}.\arabic{definition}}
        \renewcommand{\thecorollary}{\Alph{appendixc}.\arabic{corollary}}
        \renewcommand{\theequation}{\Alph{appendixc}.\arabic{equation}}
        \noindent{\tenbf Appendix \theappendixc #1}\par\vspace{5pt}}
\newcommand{\subappendix}[1] {\vspace{12pt}
        \refstepcounter{subappendixc}
        \noindent{\bf Appendix \thesubappendixc. {\kern1pt \bfit #1}}
        \par\vspace{5pt}}
\newcommand{\subsubappendix}[1] {\vspace{12pt}
        \refstepcounter{subsubappendixc}
        \noindent{\rm Appendix \thesubsubappendixc. {\kern1pt \tenit #1}}
        \par\vspace{5pt}}

\newcommand{\textlineskip}{\baselineskip=13pt}
\newcommand{\smalllineskip}{\baselineskip=10pt}

\def\eightcirc{
\begin{picture}(0,0)
\put(4.4,1.8){\circle{6.5}}
\end{picture}}
\def\eightcopyright{\eightcirc\kern2.7pt\hbox{\eightrm c}}

\def\abstracts#1#2#3{{
        \centering{\begin{minipage}{4.5in}\baselineskip=10pt\footnotesize
        \parindent=0pt #1\par
        \parindent=15pt #2\par
        \parindent=15pt #3
        \end{minipage}}\par}}

\renewenvironment{thebibliography}[1]
        {\frenchspacing
         \ninerm\baselineskip=11pt
         \begin{list}{\arabic{enumi}.}
        {\usecounter{enumi}\setlength{\parsep}{0pt}
         \setlength{\leftmargin 12.7pt}{\rightmargin 0pt} 
         \setlength{\itemsep}{0pt} \settowidth
        {\labelwidth}{#1.}\sloppy}}{\end{list}}

\newcounter{itemlistc}
\newcounter{romanlistc}
\newcounter{alphlistc}
\newcounter{arabiclistc}

\newcommand{\fcaption}[1]{
        \refstepcounter{figure}
        \setbox\@tempboxa = \hbox{\footnotesize Fig.~\thefigure. #1}
        \ifdim \wd\@tempboxa > 5in
           {\begin{center}
        \parbox{5in}{\footnotesize\smalllineskip Fig.~\thefigure. #1}
            \end{center}}
        \else
             {\begin{center}
             {\footnotesize Fig.~\thefigure. #1}
              \end{center}}
        \fi}

\newcommand{\tcaption}[1]{
        \refstepcounter{table}
        \setbox\@tempboxa = \hbox{\footnotesize Table~\thetable. #1}
        \ifdim \wd\@tempboxa > 5in
           {\begin{center}
        \parbox{5in}{\footnotesize\smalllineskip Table~\thetable. #1}
            \end{center}}
        \else
             {\begin{center}
             {\footnotesize Table~\thetable. #1}
              \end{center}}
        \fi}

\def\@citex[#1]#2{\if@filesw\immediate\write\@auxout
        {\string\citation{#2}}\fi
\def\@citea{}\@cite{\@for\@citeb:=#2\do
        {\@citea\def\@citea{,}\@ifundefined
        {b@\@citeb}{{\bf ?}\@warning
        {Citation `\@citeb' on page \thepage \space undefined}}
        {\csname b@\@citeb\endcsname}}}{#1}}

\newif\if@cghi

\def\pmb#1{\setbox0=\hbox{#1}
        \kern-.025em\copy0\kern-\wd0
        \kern.05em\copy0\kern-\wd0
        \kern-.025em\raise.0433em\box0}

\def\fnt#1#2{\footnotetext{\kern-.3em
        {$^{\mbox{\scriptsize #1}}$}{#2}}}

\def\fpage#1{\begingroup
\voffset=.3in
\thispagestyle{empty}\begin{table}[b]\centerline{\footnotesize #1}
        \end{table}\endgroup}

\def\runninghead#1#2{\pagestyle{myheadings}
\markboth{{\protect\footnotesize\it{\quad #1}}\hfill}
{\hfill{\protect\footnotesize\it{#2\quad}}}}
\font\tenrm=cmr10
\font\tenit=cmti10
\font\tenbf=cmbx10
\font\bfit=cmbxti10 at 10pt
\font\ninerm=cmr9

\font\eightrm=cmr8

\def\qed{\hbox{${\vcenter{\vbox{                        
   \hrule height 0.4pt\hbox{\vrule width 0.4pt height 6pt
   \kern5pt\vrule width 0.4pt}\hrule height 0.4pt}}}$}}

\input{psfig.tex}
\input{epsfig.sty}

\newcommand{\AmS}{{\protect\the\textfont2
  A\kern-.1667em\lower.5ex\hbox{M}\kern-.125emS}}

\newcommand\ApJ{{\sl Astrophys.\ J.\ }}

\newcommand\PRD{{\sl Phys.\ Rev.\ D\ }}
\newcommand\PRL{{{\sl Phys. Rev.\ Lett.\ }}}

\newcommand\PLB{{\sl Phys.\ Lett.\ B\ }}

\def\fun#1#2{\lower3.6pt\vbox{\baselineskip0pt\lineskip.9pt
  \ialign{$\mathsurround=0pt#1\hfil##\hfil$\crcr#2\crcr\sim\crcr}}}
\def\lesssim{\mathrel{\mathpalette\fun <}}
\def\gtrsim{\mathrel{\mathpalette\fun >}}

\long\def\comment#1{}

\def\be{\begin{equation}}
\def\ee{\end{equation}}
\def\bea{\begin{eqnarray}}
\def\eea{\end{eqnarray}}

\def\VEV#1{\left\langle #1\right\rangle}
\def\sec{\ifmmode \,\, {\rm sec} \else sec \fi}
\def\eV {\ifmmode \,\, {\rm eV} \else eV \fi}
\def\keV{\ifmmode \,\, {\rm keV} \else keV \fi}
\def\MeV{\ifmmode \,\, {\rm MeV} \else MeV \fi}
\def\GeV{\ifmmode \,\, {\rm GeV} \else GeV \fi}
\def\TeV{\ifmmode \,\, {\rm TeV} \else TeV \fi}
\def\fm{\ifmmode \,\, {\rm fm} \else TeV \fi}
\def\pbarn{\ifmmode \,\, {\rm pb} \else pb \fi}
\def\km{\ifmmode {\rm km}\, \else km \fi}
\def\Mpc{\ifmmode {\rm Mpc}\, \else Mpc \fi}
\def\Gyr{\ifmmode {\rm Gyr}\, \else Gyr \fi}

\def\fun#1#2{\lower3.6pt\vbox{\baselineskip0pt\lineskip.9pt
  \ialign{$\mathsurround=0pt#1\hfil##\hfil$\crcr#2\crcr\sim\crcr}}}
\def\la{\mathrel{\mathpalette\fun <}}
\def\ga{\mathrel{\mathpalette\fun >}}

\def\etal{{\it et al.}}

\def\sbar#1{\kern 0.8pt
        \overline{\kern -0.8pt #1 \kern -0.8pt}
        \kern 0.8pt}  

\def\meter{\ifmmode \,\, {\rm m} \else m \fi}
\def\yr {\ifmmode \,\, {\rm yr} \else yr \fi}

\def\sr{\ifmmode \,\, {\rm sr} \else sr \fi}

\def\hatn{{\bf \hat n}}

\def\slashchar#1{\setbox0=\hbox{$#1$}           
   \dimen0=\wd0                                 
   \setbox1=\hbox{/} \dimen1=\wd1               
   \ifdim\dimen0>\dimen1                        
      \rlap{\hbox to \dimen0{\hfil/\hfil}}      
      #1                                        
   \else                                        
      \rlap{\hbox to \dimen1{\hfil$#1$\hfil}}   
      /                                         
   \fi}

\begin{document}
\runninghead{Dark Matter and Dark Energy}{Dark Matter and Dark Energy}

\normalsize\textlineskip
\thispagestyle{empty}
\setcounter{page}{1}


\vspace*{0.88truein}

\fpage{1}
\centerline{\Large\bf DARK MATTER AND DARK
ENERGY\footnote{Submitted for publication in ``Visions of
Discovery'' (in honor of Charles Townes), to be published by
Cambridge University Press.}}
\vspace*{0.37truein}
\centerline{\large MARC KAMIONKOWSKI\footnote{kamion@tapir.caltech.edu}}
\vspace*{0.015truein}
\centerline{\it California Institute of Technology, Mail Code
130-33, Pasadena, CA~~91125, USA}

\vspace*{0.21truein}
\abstracts{
This is a short review, aimed at a general audience, of several
current subjects of research in cosmology.  The
topics discussed include the cosmic microwave background (CMB),
with particular emphasis on its relevance for testing inflation;
dark matter, with a brief review of astrophysical evidence and
more emphasis on particle candidates; and cosmic acceleration
and some of the ideas that have been put forward to explain it.  A
glossary of technical terms and acronyms is provided.}{}{}


\vspace*{1pt}\textlineskip      


\section{Introduction}

Now is the time to be a cosmologist.
We have obtained through remarkable
technological advances and heroic and ingenious experimental
efforts a direct and extraordinarily detailed picture of the
early Universe and maps of the distribution of matter on the
largest scales in the Universe today.  We have, moreover, an
elegant and precisely quantitative physical model for the origin
and evolution of the Universe.  However, the model invokes new
physics, beyond the standard model plus general relativity, not
just once, but at least thrice:  (1) Inflation, the
physical mechanism for making the early Universe look precisely
as it does, posits some new ultra-high-energy physics; we don't
know, however, what it is.  (2) The growth of
large-scale-structure and the dynamics of galaxies and galaxy
clusters requires that we invoke the existence of collisionless
particles or objects; we don't know what this stuff is.  (3) The
accelerated expansion of the Universe
requires the introduction of a new term, of embarrassingly small
value, in Einstein's equation, a modification of general
relativity, and/or the introduction of some negative-pressure
``dark energy,'' again, the nature of which remains a mystery.

In science, though, confusion and uncertainty are opportunity.
There are well-defined but fundamental questions to be answered
and data arriving to guide theory.  Ongoing and forthcoming observations
and experiments will in the next few years provide empirical
information about the new physics responsible for inflation, the
nature of the dark matter, and the puzzle of accelerated
expansion.  Future discoveries may help us understand the new physics
that unifies the strong, weak, and
electromagnetic interactions, as well as gravity.  There are also
always the prospects for a major paradigm shift in physics,
which may be required to unify gravity with quantum mechanics.

In this Chapter, I review the current status of our
cosmological model as well as its shortcomings and the questions
it leaves unanswered, and I discuss possible answers to these
questions and possible avenues towards testing these answers.
In particular, I focus on three subjects.  In the next Section,
I discuss the cosmic microwave background and inflation.
Although the main subject of this review is dark matter and dark
energy, the paradigm upon which many of our
observations---including those that suggest dark matter and dark
energy---are interpreted is a Universe with primordial
perturbations remarkably like those predicted by inflation.
Moreover, the most precise information we have now about the
Universe and its contents is the cosmic microwave background,
and so it behooves us to review this subject before considering
dark matter and dark energy.  I then move on in Section 3 to
dark matter.  I focus primarily on particle dark matter and
discuss the prospects for detection of such dark matter, as well
as some variations on the simplest particle models for dark
matter.  Section 4 reviews the cosmic-acceleration puzzle.  I
review the evidence and then discuss several possible
solutions.  Section 5 provides some closing remarks, and Section
6 contains a glossary (prepared in collaboration with Adrian
Lee) of technical terms and acronyms used in this review and in
the Chapter in this volume by Adrian Lee.

\section{The Cosmic Microwave Background and Inflation}

A confluence of theoretical developments and technological
breakthroughs during the past decade have transformed the cosmic 
microwave background (CMB) into a precise tool for determining the
contents, largest-scale structure, and origin of the Universe.
Tiny (few parts in $10^5$) angular variations in the temperature
of the CMB were discovered in the early 1990s by the
Differential Microwave Radiometer (DMR) aboard NASA's Cosmic
Background Explorer ({\sl COBE}) \cite{Smoetal90}, and during
the past few years, high--signal-to-noise
high--angular-resolution ($\sim0.2^\circ$) CMB temperature maps
have been obtained \cite{smallscaleCMB}.  These provide the very 
first snapshots of the Universe as it was roughly 380,000
years after the big bang, nearly 14 billion years ago, when 
electrons and light nuclei first combined to form neutral
hydrogen and helium.

These new maps have provided several extraordinary
breakthroughs.  The most striking among these is fairly robust
evidence that the Universe is flat and that large-scale
structure (galaxies, clusters of galaxies, and even larger
structures) grew via gravitational infall from a nearly
scale-invariant spectrum of primordial
density perturbations.  Both of these observations hint strongly 
that the Universe began with inflation
\cite{Gut81}, a period of accelerated
expansion in the very earliest Universe, driven by the vacuum
energy associated with some new ultra-high-energy physics.

Even more recently, the polarization of the CMB has been
detected \cite{dasipolarization} and begun to be mapped on small
scales \cite{smallscalepolarization} and detected through its
cross-correlation with the temperature \cite{WMAPTE,BoomTE}.  The
small-scale results are consistent with expectations based on
models that fit the temperature results, and the results from
three years of WMAP (Wilkinson Microwave Anisotropy Probe)
indicate that reionization likely occurred at a redshift
$z\sim10$ \cite{WMAP3polarization}.

As interesting as these results may be, the polarization may
allow even more intriguing discoveries in the future.  In
particular, a cosmological gravitational-wave background
from inflation is expected to produce a unique polarization
pattern \cite{KamKosSte97a,KamKosSte97b,SelZal97,ZalSel97}.  This
``fingerprint'' of inflation would allow us to
see directly  back to the inflationary epoch, $10^{-38}$ seconds
after the big bang!

In the following, I summarize briefly recent progress and future
prospects for CMB tests of inflation.  For a more detailed
review of the topics discussed here, see
Refs.~\cite{KamKos99,HuDod01}.

\subsection{Observation and Inflation}

Prior to the advent of these new CMB maps, the standard
hot-big-bang theory rested on the cornerstones of the expansion
of the Universe, the agreement between the observed
light-element abundances and the predictions of big-bang
nucleosynthesis (BBN), and the blackbody spectrum of the CMB.
However, this standard model still left many questions
unanswered.

\begin{figure}[htbp]
\centerline{\psfig{file=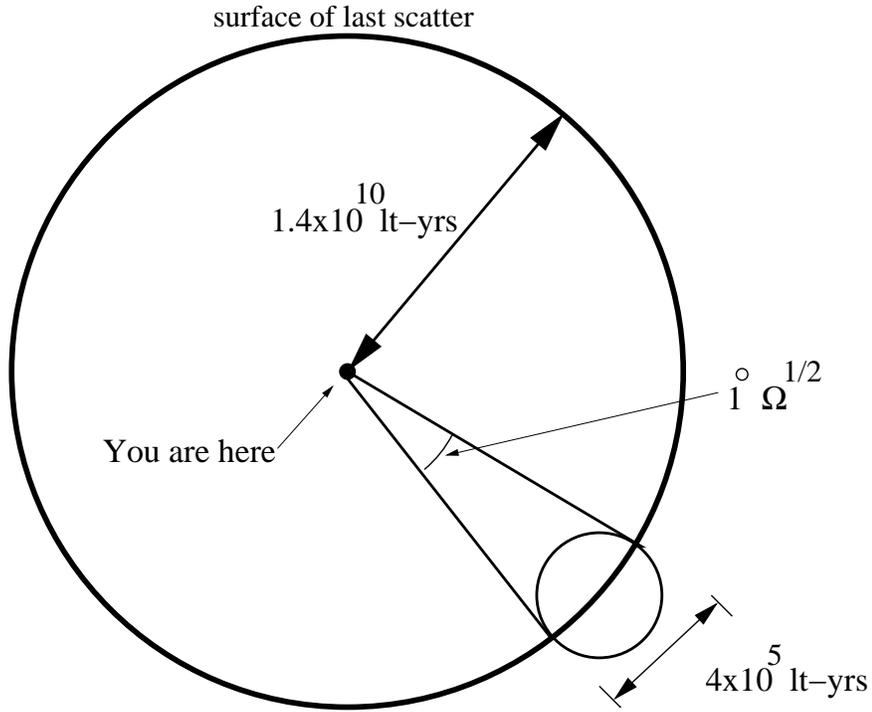,width=4.5in}}
\bigskip
\caption{The CMB that we see last scattered on a spherical
     surface roughly 14 billion light years away.  However, when
     these photons last scattered, the size of a causally
     connected region was closer to $380,000$ light years, which
     subtends an angle of roughly $1^\circ$.}
\label{fig:universe}
\end{figure}

{\it The isotropy.}  The isotropy of the CMB posed the first conundrum for the
standard big-bang theory.  The CMB photons that we see last
scattered from a spherical surface with a radius of
about 10,000 Mpc (about 14 billion light-years), when the Universe
was only about 380,000 years old, as shown in
Fig.~\ref{fig:universe}.  When these photons last
scattered, the size of a causally connected region of the
Universe was roughly 380,000 light-years, and such a region
subtends an angle of roughly one degree on the sky.  Since there
are 40,000 square degrees on the surface of the sky, {\sl COBE} was
thus looking at roughly 40,000 causally disconnected
regions of the Universe.  (Strictly speaking, {\sl COBE}'s
angular resolution was only 7 degrees, but the WMAP satellite
\cite{WMAP}, with a fraction-of-a-degree resolution saw
temperature fluctuations of no more than $\sim10^{-5}$.)  If so,
however, then why did each of these have the same temperature to
one part in $10^5$?

\begin{figure*}
\centerline{\psfig{file=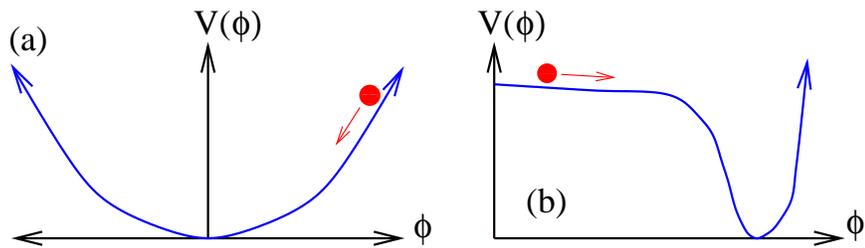,width=4.5in}}
\caption{Two toy models for the inflationary potential.}
\vskip-12pt
\label{fig:potentials}
\end{figure*}

The most appealing explanation for the isotropy
is inflation \cite{Gut81}, a period of accelerated expansion in the
very early Universe driven by the vacuum energy associated with
some ultra-high-energy phase transition.  Inflation simply
postulates some new scalar field
$\phi$ with a potential-energy density $V(\phi)$, which may
look, for example, like either of the two forms shown in
Fig.~\ref{fig:potentials}.  Suppose that at some point in the
early history of the Universe, the energy density is dominated by
the potential-energy density of this scalar field.  Then the
Friedmann equation---the general-relativistic equation that
relates the time $t$ evolution of the scale factor $a(t)$
(which quantifies, roughly speaking, the mean spacing between
galaxies) to the
energy density $\rho$---becomes $H^2\equiv(\dot
a/a)^2 \simeq 8 \pi G V/3$, where $G$ is Newton's constant (and
the dot denotes derivative with respect to time).  If
the scalar field is rolling
slowly (down a potential, like one of those shown in
Fig.~\ref{fig:potentials}), then $V$ is approximately constant
with time, and the
scale factor grows exponentially, thus blowing up a tiny
causally-connected region of the Universe into a volume large
enough to encompass the entire observable Universe.

{\it The geometry of the Universe.}
Hubble's discovery of the expansion of the Universe forced
theorists to take the general-relativistic cosmological models
of Einstein, de Sitter, Lemaitre, Friedmann, Robertson, and
Walker seriously.  These models showed that
the Universe must be open, closed, or flat.  A flat Universe is
one in which the three spatial dimensions satisfy the laws of
Euclidean geometry; in a closed Universe, the laws of
geometry for the three spatial dimensions resemble those for a
three-dimensional analogue of the surface of a sphere; and an
open Universe is a three-dimensional analogue of the surface of
a saddle.  In a (flat,closed,open) Universe, the interior angles 
of a triangle sum to ($180^\circ$,$>180^\circ$,$<180^\circ$),
the circumference of a circle is ($2\pi$,$<2\pi$,$>2\pi$) times
its radius, and (most importantly) the angular size of an object
of physical size $l$ observed at a distance $d$ is
($\theta=l/d$,$\theta>l/d$,$\theta<l/d$).
General relativity dictates that the geometry is related to
$\Omega_{\rm tot}\equiv\rho_{\rm tot}/\rho_c$, the {\it total} density
$\rho_{\rm tot}$ of the Universe in units of the critical
density $\rho_c\equiv 3 H_0^2/8\pi G$, where $H_0$ is the
expansion rate today.  A value of $\Omega_{\rm tot}>1$,
$\Omega_{\rm tot}=1$, and $\Omega_{\rm tot}<1$ corresponds
respectively to a closed, flat, and open universe. For 70 years
after Hubble's discovery, measurements of $\Omega_{\rm tot}$
were unable to achieve the precision required to determine the
geometry.

However, the high-sensitivity high-angular-resolution maps of
the CMB temperature that have now been obtained have allowed a
direct test of the geometry \cite{KamSpeSug94}.  These
experiments have measured the temperature $T(\hatn)$ as a
function of position $\hatn$ on the sky.  The coefficients in a
spherical-harmonic expansion of $T(\hatn)$ are
\begin{equation}
     a_{(\ell m)}^{\rm T} = \int\, d^2\hatn \, T(\hatn) \, Y_{(\ell m)}(\hatn),
\end{equation}
and from them we can construct a
power spectrum, $C_\ell = \VEV{ |a_{lm}|^2}$, where the
average is over all $2\ell+1$ values of $m$.

\begin{figure*}[htbp]
\centerline{\psfig{file=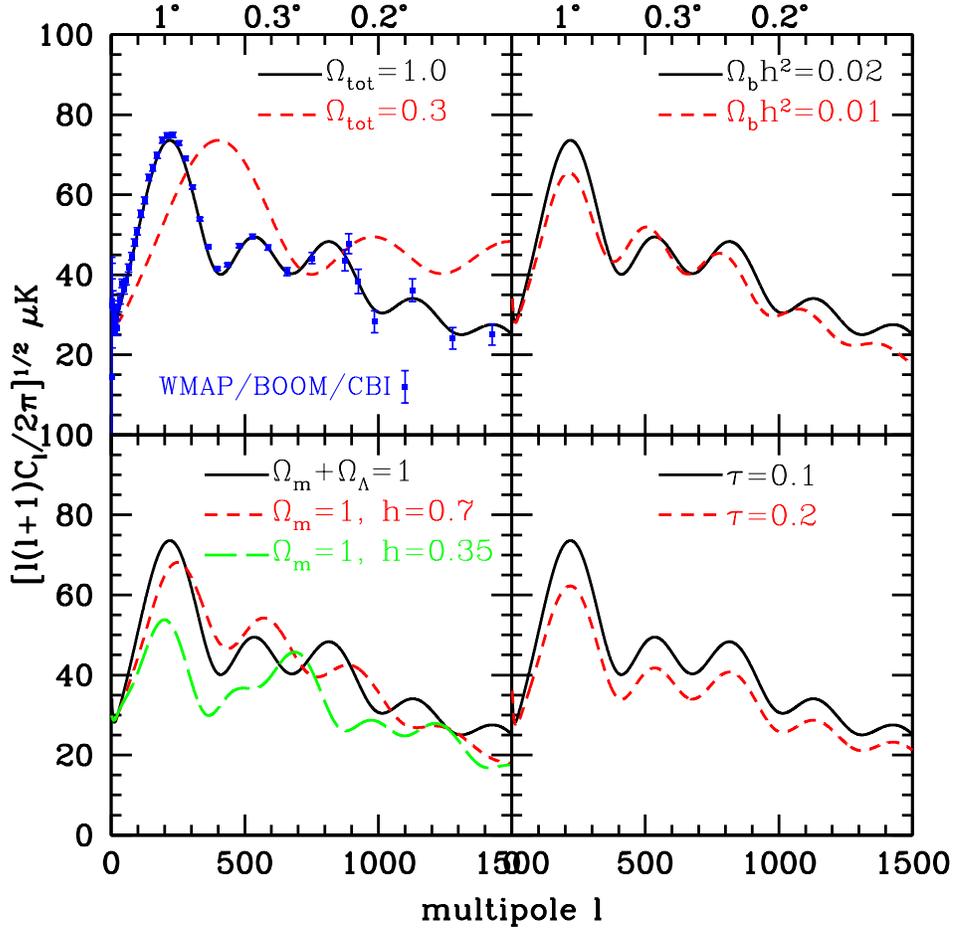,width=5.0in}}
\caption{CMB power spectra.  The solid curve in each panel show
          the current best-fit model, with 
          $\{\Omega_m h^2,\Omega_b
          h^2,h,n_s,\tau\}=\{0.1277,0.02229,0.732,0.958,0.089\}$
          \cite{Spergel06}.  To indicate the precision of
          current experiments, data points from WMAP (small
          $\ell$), BOOMERanG (intermediate $\ell$), and CBI
          (large $\ell$) are shown.
          Each panel shows the effect of independent variation
          of a single cosmological parameter.  The Planck
          satellite, to be launched in 2008, should have error
          bars from $\ell=2$ to $\ell=1500$ (and higher) that
          are no thicker than the thickness of the curve.}
\label{fig:models}
\end{figure*}

Given a structure-formation theory (e.g., inflation)
as well as the values of the
cosmological parameters, it is straightforward to predict the
CMB power spectrum.  Such calculations take into account the
evolution of density perturbations as governed by Einstein's
equations as well as the motion and distributions of baryons,
dark matter, neutrinos, and photons in these perturbations
as governed by their fluid and Boltzmann equations.  The solid
curves in Fig.~\ref{fig:models} show results of such
calculations for inflationary density perturbations with a set
of cosmological parameters consistent with current data:
a flat ($\Omega_m+\Omega_\Lambda=1$) model with 
$\{\Omega_m
h^2,\Omega_b
h^2,h,n_s,\tau\}=\{0.1277,0.02229,0.732,0.958,0.089\}$
\cite{Spergel06}.  Each panel 
shows the effect of independent variation of one of the
cosmological parameters.  The acoustic-peak structure, first
predicted by Sunyaev and Zeldovich \cite{SunZel70} and Peebles
and Yu \cite{PeeYu70}, is due to the propagation of density
perturbations as acoustic waves in the primordial plasma.  As
illustrated, the height, width, and spacing of the acoustic
peaks in the angular spectrum depend on these (and other)
cosmological parameters.

In particular, the location of the first peak is determined by
the angle subtended by the acoustic horizon at the surface of
last scatter.  This is $\theta\simeq1^\circ$ in a flat Universe,
and it scales roughly as $\Omega_{\rm tot}^{1/2}$ in a non-flat
Universe for the geometric reasons discussed above.  Thus, the
first peak should be located at $\ell\sim220 \Omega_{\rm tot}^{-1/2}$
\cite{KamSpeSug94,Junetal96a}.  As of 2000, balloon data already
suggested $\Omega_{\rm tot}=1.11\pm0.07^{+0.13}_{-0.12}$
(statistical and systematic errors), and WMAP
now constrains $\Omega_{\rm tot}=1.02\pm0.02$ \cite{Spergel06}.

Thus, a new question arises: i.e., why is the Universe flat?
An answer to this also comes from inflation.  If inflation is to 
last sufficiently long to explain the isotropy problem, then it
must produce a flat Universe.  This can be seen from the
form of the Friedmann equation,
\begin{equation}
     H^2 \equiv \left( \frac{\dot a}{a}\right)^2 = \frac{8\pi G
      V}{3} - \frac{k}{a^2},
\end{equation}
during inflation.  After inflation sets in, $a\propto e^{Ht}$,
$V\sim$constant, and so the curvature term $k/a^2 \propto
a^{-2Ht}$ decays exponentially.

{\it The origin of large-scale structure.}
Another fundamental aim of modern cosmology is to understand the
origin of galaxies, clusters of galaxies, and structures on even 
larger scales.
The simplest and most plausible explanation---that these mass
inhomogeneities grew from tiny density perturbations in the
early Universe via gravitational instability---was confirmed by
the tiny temperature fluctuations seen in {\sl COBE}
\cite{Smoetal90}.  These temperature fluctuations 
are due to density perturbations at the surface of last scatter;
photons from denser regions climb out of deeper potential wells
and thus appear redder than those from underdense regions
\cite{SacWol67}.  The observed temperature-fluctuation
amplitude is in good agreement with the density-perturbation
amplitude required to seed large-scale structure.

But this gives rise to yet another question: where did these
primordial perturbations come from?  Before {\sl COBE}, there was no
shortage of ideas: perturbations may have come from (just to
list some names) inflation,
late-time phase transitions, a loitering Universe, scalar-field
ordering, topological defects (such as cosmic strings, domain
walls, textures, or global monopoles), superconducting cosmic
strings, a Peccei-Quinn symmetry-breaking transition, etc.
However, after {\sl COBE}, density perturbations like those produced
by inflation \cite{perturbations} became the frontrunners; in
particular, models with anything other than primordial adiabatic
perturbations generically predict more large-angle temperature
fluctuations than models with adiabatic perturbations \cite{JafSteFri94}.
Now, with the CMB maps obtained the last seven years, any
alternatives to inflationary perturbations
have become increasingly difficult to reconcile with the
data, and the detailed acoustic-peak structure in the CMB power
spectra are in beautiful agreement with inflationary models. 
The CMB shows that primordial
perturbations were nearly scale invariant, and extend to
distance scales that were larger than the horizon at
the surface of last scatter.  These superhorizon perturbations
are another feather in inflation's cap.

\subsection{What is the New Physics Responsible for Inflaton?}

The agreement between inflation's predictions and the data obtained so
far suggest that we may be on the right track with inflation,
and this motivates us to consider new, more precise, tests and
to think more deeply about the physics
of inflation.  Although the idea behind inflation
is simple, we do not know what new physics is responsible for
inflation.  Another way to ask this question is when, in the
early history of the Universe, did inflation occur?  Since the
temperature of the Universe increases monotonically as we go to
earlier cosmological times, we may also ask at what temperature
did inflation occur?  To first get our bearings, we note that
the Universe is today about 14 billion years old, and the
temperature is 2.7 K, corresponding to a typical thermal energy
of $10^{-3}$ eV, small compared even with molecular-transition
energies.  Stars and galaxies formed several billion years after
the big bang.  Electrons and protons first
combined to form hydrogen atoms roughly 380,000 years after the big
bang, at a temperature of roughly 3000 K, when the mean thermal
energies of the CMB were comparable to the ionization energy for
the hydrogen atom.  CMB photons also decoupled from the
primordial plasma at about this time (as the
free electrons from which they scattered disappeared).  Neutrons 
and protons were first assembled into light nuclei (D, $^3$He,
$^4$He, $^7$Li) a few seconds to minutes after the big bang,
when the CMB thermal energies fell below an MeV, the binding energy
per nucleon.  Quarks presumably collected into hadrons at a
temperature of roughly 100 MeV, although the details are still
unclear.


To extrapolate further back in time, we need to understand the
physics of elementary particles at higher energies.  We now have 
a secure model that unifies the electromagnetic and weak
interactions at energies $\sim100$ GeV.  This electroweak
symmetry would have first been broken at a cosmological
electroweak phase transition roughly $10^{-9}$ seconds after the 
big bang.  Similarities between the mathematical structure of
the strong and electroweak interactions have led particle
theorists to postulate a grand unified theory (GUT) that would
be first broken at an energy $\sim10^{16}$ GeV, roughly
$10^{-38}$ seconds after the big bang.  String theories go even
further and provide a mechanism for incorporating the strong, weak,
and electromagnetic interactions into a quantum theory of gravity at the
Planck scale, $10^{19}$ GeV.  There are also other interesting
ideas in particle theory, such as Peccei-Quinn symmetry (a new
symmetry postulated in order to solve the strong-CP problem; see
Section 1.3.9), which
would be broken at $\sim10^{12}$ GeV and supersymmetry
(postulated in order to explain the hierarchy between the GUT
scale and the electroweak scale), which must have also been
broken at some point.

Inflation was originally conceived in association with grand
unification, and many (although not all) theorists would still
consider GUTs to provide the most natural home for inflation.
However, the ingredients necessary for inflation may also be
found in string theories, Peccei-Quinn symmetry breaking,
supersymmetry breaking, or even at the electroweak scale.  In
recent years, a vast array of inflationary models with extra
dimensions have been explored (see, e.g., Ref.~\cite{jmr}).
Ref. \cite{LytRio99} reviews particle-physics models of inflation.

\subsection{Inflation and CMB Polarization}

One way to determine the new physics responsible for inflation
is to ask, what is the height
$V$ of the inflaton potential? or equivalently, what is the
energy scale $E_{\rm infl}$, defined by $V=E_{\rm infl}^4$, of inflation?
If inflation had something to do with grand unification, then we
might expect $E_{\rm infl}\sim 10^{15-16} \,  {\rm GeV}$; if it
had to do with some lower-energy physics, then $E_{\rm infl}$
should be correspondingly lower (e.g., Peccei-Quinn symmetry
breaking would suggest $E_{\rm infl}\sim10^{12}$ GeV).

\begin{figure}[htbp]
\centerline{\epsfig{figure=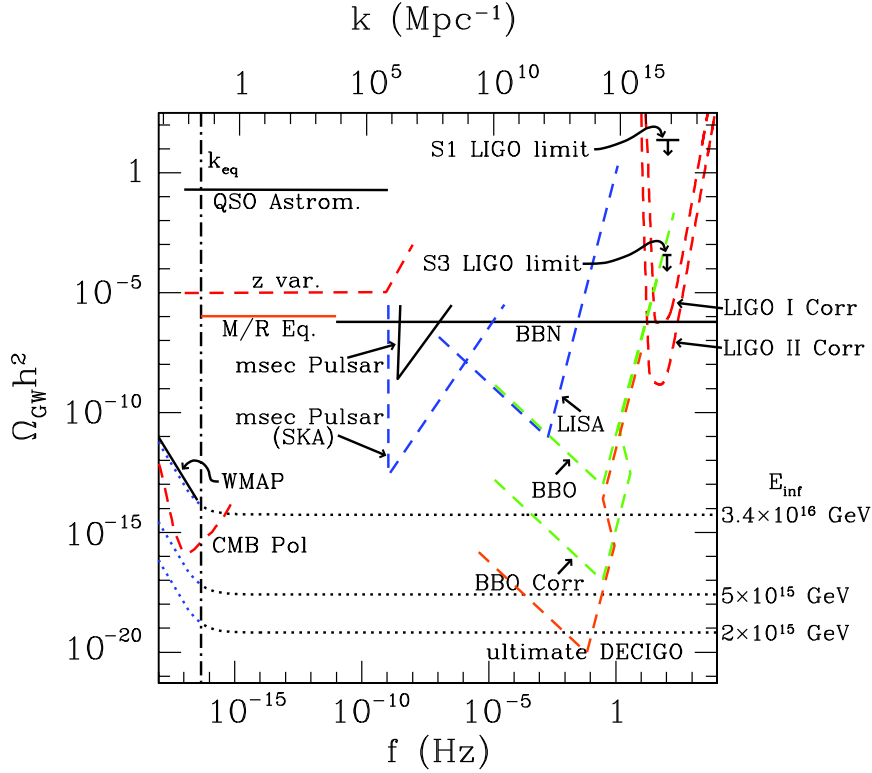,width=4.5in}}
\medskip
\caption{Current limits and projected sensitivities to a
     stochastic gravitational-wave background versus the
     gravitational-wave frequency.  The solid curves all
     indicate current upper limits, while the various broken curves
     indicate projected sensitivities. The ``M/R'' line comes
     CMB constraints to the epoch of matter-radiation
     equality \protect\cite{SmiPieKam06}.
     Curves corresponding to scale-invariant (i.e., $n_t=0$)
     gravitational-wave backgrounds are shown (dotted curves),
     labeled by the associated inflationary energy scales.
     The amplitude of CMB temperature fluctuations currently
     constrains this value to be below
     $3.36 \times 10^{16}$ GeV, but only at frequencies
     $f<10^{-16}$ Hz.  Future CMB
     measurements may be able to reach energy scales near
     $10^{15}$ GeV at these frequencies.  The ``QSO Astrom''
     curve is a limit from
     quasar astrometry, and the ``z var'' is a forecast for
     future redshift measurements.  The S1 and S3 points are
     upper limits from the Laser Interferometric Gravitational
     Wave Observatory (LIGO) \protect\cite{LIGO} and the other curves
     are forecasts for future LIGO sensitivities.  The LISA
     curve shows forecasts for the future NASA/ESO Laser
     Interferometric Space Observatory and the BBO and DECIGO
     curves show forecasts for sensitivities for two space-based
     observatories now under study (the ``Corr'' designation is
     for a configuration in which the signals from two detectors
     or detector arrays are correlated against one
     another---e.g., for LIGO, if the signals from the Hanford
     and 
     Louisiana sites are correlated).  The two ``pulsar'' curves
     show current and future (from the Square Kilometer Array;
     SKA) sensitivities from pulsar timing.
     The WMAP and ``CMB Pol'' curves show the current upper
     limit from WMAP and the sensitivity forecast for CMBPol, a
     satellite mission now under study.  From
     Ref.~\protect\cite{SmiKamCoo05}.}
\label{fig:spectrum}
\end{figure}

The energy scale of inflation can be determined with the gravitational-wave
background.  Through quantum-mechanical effects analogous to the
production of Hawking radiation from black holes, inflation
produces a stochastic cosmological background
of gravitational waves \cite{AbbWis84}.  It is well 
known that the temperature of the Hawking radiation emitted from
a (non-charged and non-spinning) black hole is determined
exclusively by the black-hole mass, as this determines the
spacetime curvature around the black hole.  Likewise, during
inflation, the spacetime curvature is determined exclusively by
the cosmological energy density, which is just the inflaton-potential height
$V=E_{\rm infl}^4$ during inflation.
Calculation shows that the amplitude of the gravitational-wave
background is proportional to $(E_{\rm infl}/m_{\rm Pl})^2$, where 
$m_{\rm Pl}\simeq 10^{19}$ GeV is the Planck mass.  Therefore,
if we can detect this gravitational-wave background and
determine its amplitude, we learn the energy scale of inflation
and thus infer the new physics responsible for inflation.
Fig.~\ref{fig:spectrum} shows the amplitude of the
gravitational-wave background, as a function of frequency, from
simple inflation models that produce a
scale-invariant spectrum, one with a spectral index $n_t=0$
(where $n_t$ measures the relative amplitude of short- versus
long-wavelength gravitational waves, and the subscript ``t''
stands for tensor perturbations, another term for gravitational
waves) for several different $E_{\rm infl}$.  More generally,
inflation models usually predict $n_t<0$, implying less power on
smaller scales (or larger frequencies).  The Figure also
shows current constraints and future prospects for detection, as
we now discuss.

\begin{figure}[htbp]
\centerline{\psfig{file=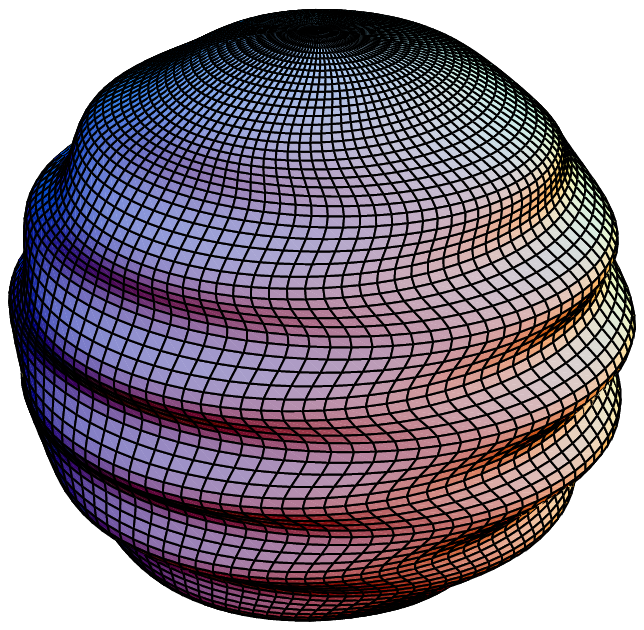,width=4.5in}}
\bigskip
\caption{The shape of the surface of last scatter if a single
         gravitational wave propagates in the vertical direction
         through the Universe.}
\label{fig:wave}
\end{figure}

\begin{figure}[htbp]
\centerline{\psfig{file=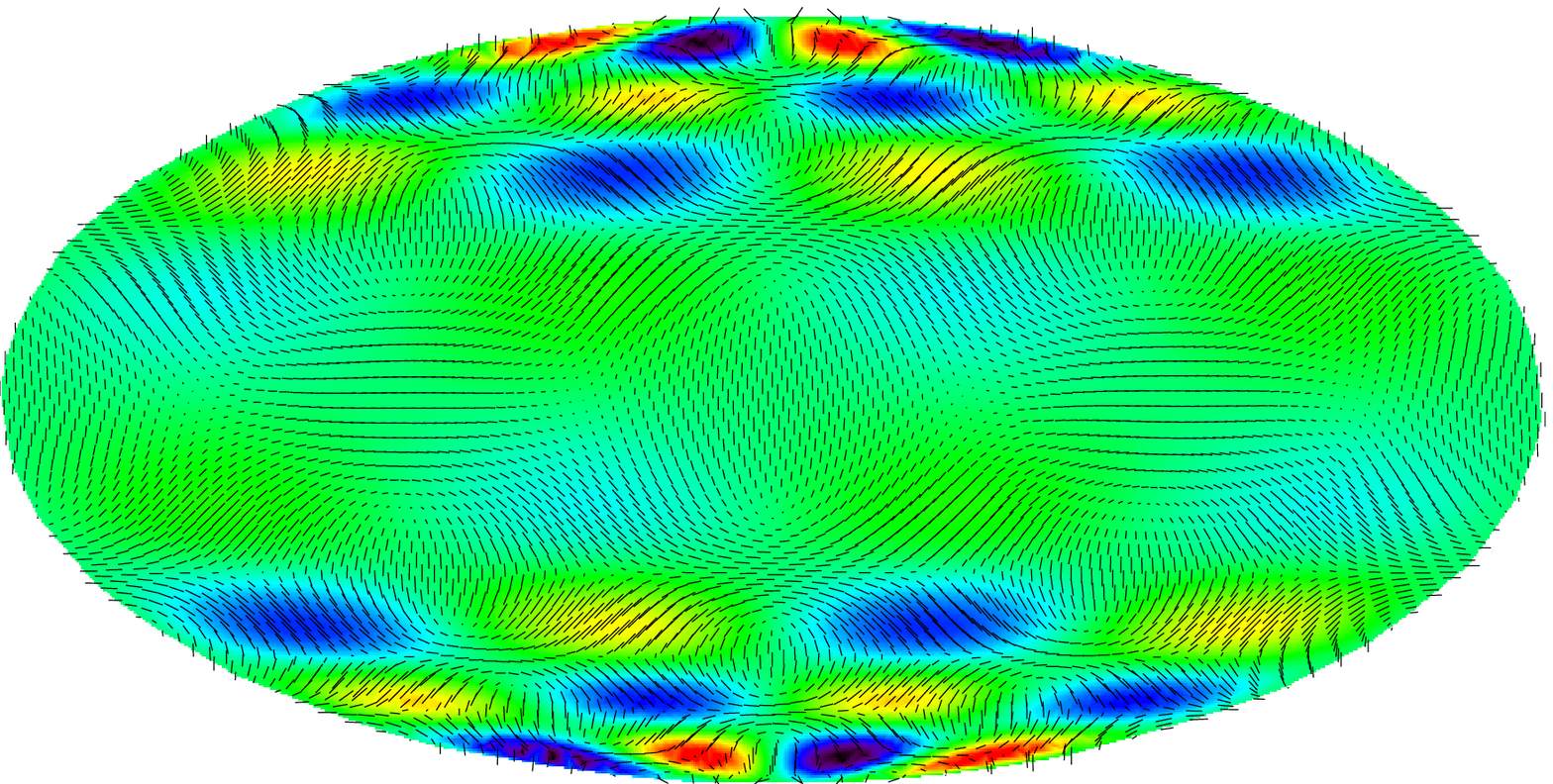,width=4.5in}}
\bigskip
\caption{The CMB temperature and polarization pattern induced by 
         a single gravitational wave.  This is an equal-area
     representation of the full spherical surface of the sky.
     If this were a map of the Earth, North America, South
     America, Australia, and Eurasia would occupy, respectively,
     the upper-left, lower-left, lower-right, and upper-right
     quadrants.  The orientation of the
     lines reflects that of the polarization, and the size is
     proportional to the polarization amplitude.  The gray scale
     represents temperature fluctuations that span one part in
     $10^5$.  The quadrupolar variation of the
     temperature/polarization pattern can be seen as one travels
     along a curve of constant latitude, and the wavelike
     pattern can be seen as one moves along a constant
     longitude.  From Ref. \protect\cite{CalKamWad99}.}
\label{fig:cmbwave}
\end{figure}

Perhaps the most promising avenue toward detecting the
inflationary-gravitational-wave (IGW) background is with
the CMB, at ultra-low gravitational-wave frequencies,
gravitational waves with wavelengths  comparable to the
observable Universe.  Just as an electromagnetic wave is
detected through observation
of the motion its oscillating electromagnetic fields induce in
test charges, a gravitational wave is detected through the
motion that its oscillating gravitational field induces in test 
masses.  More precisely, a gravitational plane wave will induce
a quadrupolar oscillation in a ring of test masses located in a
plane perpendicular to the wave's direction of propagation.  Now 
suppose a long-wavelength gravitational wave is propagating
through the Universe.  Then the primordial plasma from which the
CMB photons we observe last scatter can be used as a sphere of
test masses.  The gravitational wave will induce motions in this 
primordial plasma, as shown in Fig.~\ref{fig:wave}.  If photons
last scatter from plasma that is moving away from or toward us,
then the photons will appear red- or blue-shifted.  Thus, that
single gravitational wave will induce a temperature pattern on
the CMB sky that looks like that shown in
Fig.~\ref{fig:cmbwave}.  Hence, the WMAP limit to $\Omega_{\rm
gw}h^2$ shown on the left-hand side of Fig.~\ref{fig:spectrum}.

\begin{figure}[htbp]
\centerline{\psfig{file=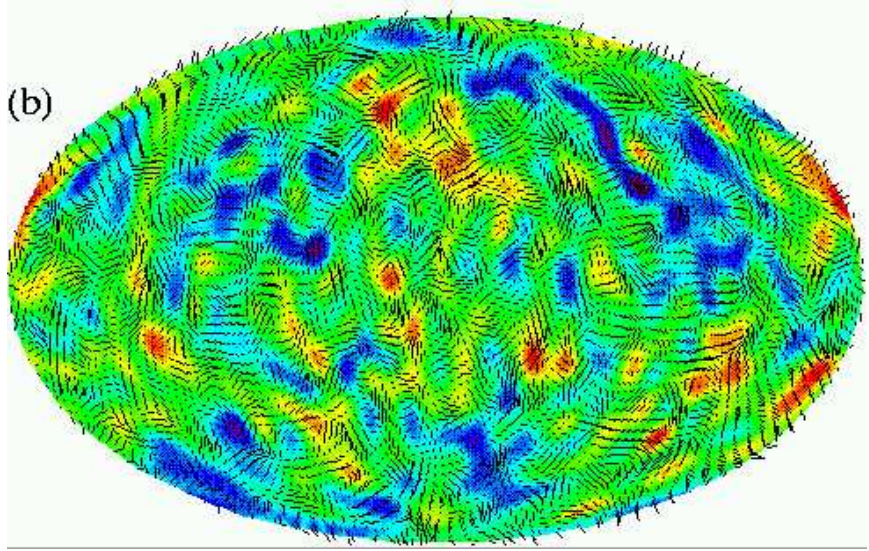,width=4.5in}}
\bigskip
\caption{Simulated CMB temperature/polarization pattern induced by
     inflationary gravitational waves.  From
     Ref.~\protect\cite{CalKamWad99}.}
\label{fig:two}
\end{figure}

Inflation predicts a {\it stochastic}
background of such gravitational waves, rather than a single
gravitational wave, so the sky should look more like
Fig.~\ref{fig:two}.  However, a plausible spectrum of density
perturbations could produce a temperature map that looks almost
identical.  More precisely, gravitational waves would produce
temperature fluctuations only on large angular scales, so their
presence would increase the power at $\ell \lesssim 50$ relative 
to the power in the peaks at $\ell \gtrsim100$.  However,
re-scattering of some CMB photons from electrons that would have 
been reionized during the production of the first stars and
quasars would reduce the power in the peaks relative to that at
large angles, thus mimicking the effect of gravitational waves
\cite{Junetal96b,Kin98,Meletal99}.

So how can we go further?  Progress can be made with the
polarization of the CMB.  A small polarization will be produced
in CMB photons because the flux of photons incident on the
electrons from which they last scatter will be anisotropic (this 
is just polarization from
right-angle scattering).  Such a polarization will be induced
for both density perturbations and gravitational waves, so
the mere detection of the polarization does not alone indicate
the presence of gravitational waves.  However, the {\it pattern}
of polarization induced on the CMB sky can be used to
distinguish gravitational waves from density perturbations.

\begin{figure}[htbp]
\centerline{\psfig{file=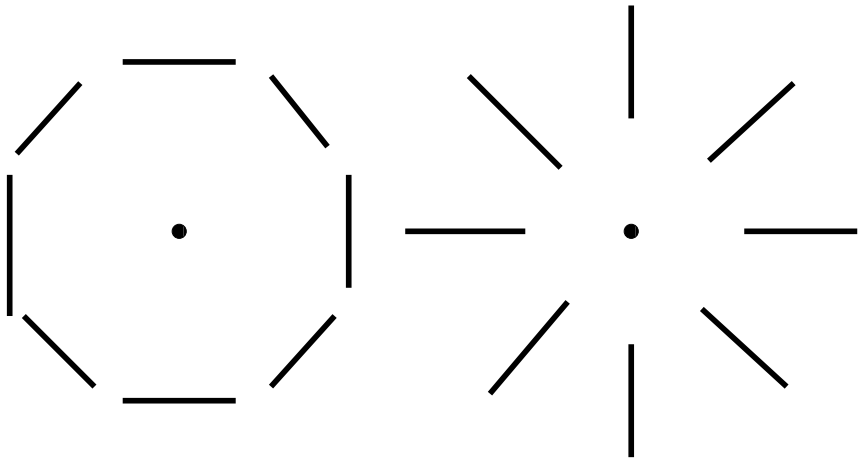,width=4.5in}}
\bigskip
\caption{Polarization pattern with no curl.}
\label{fig:sc}
\end{figure}

\begin{figure}[htbp]
\centerline{\psfig{file=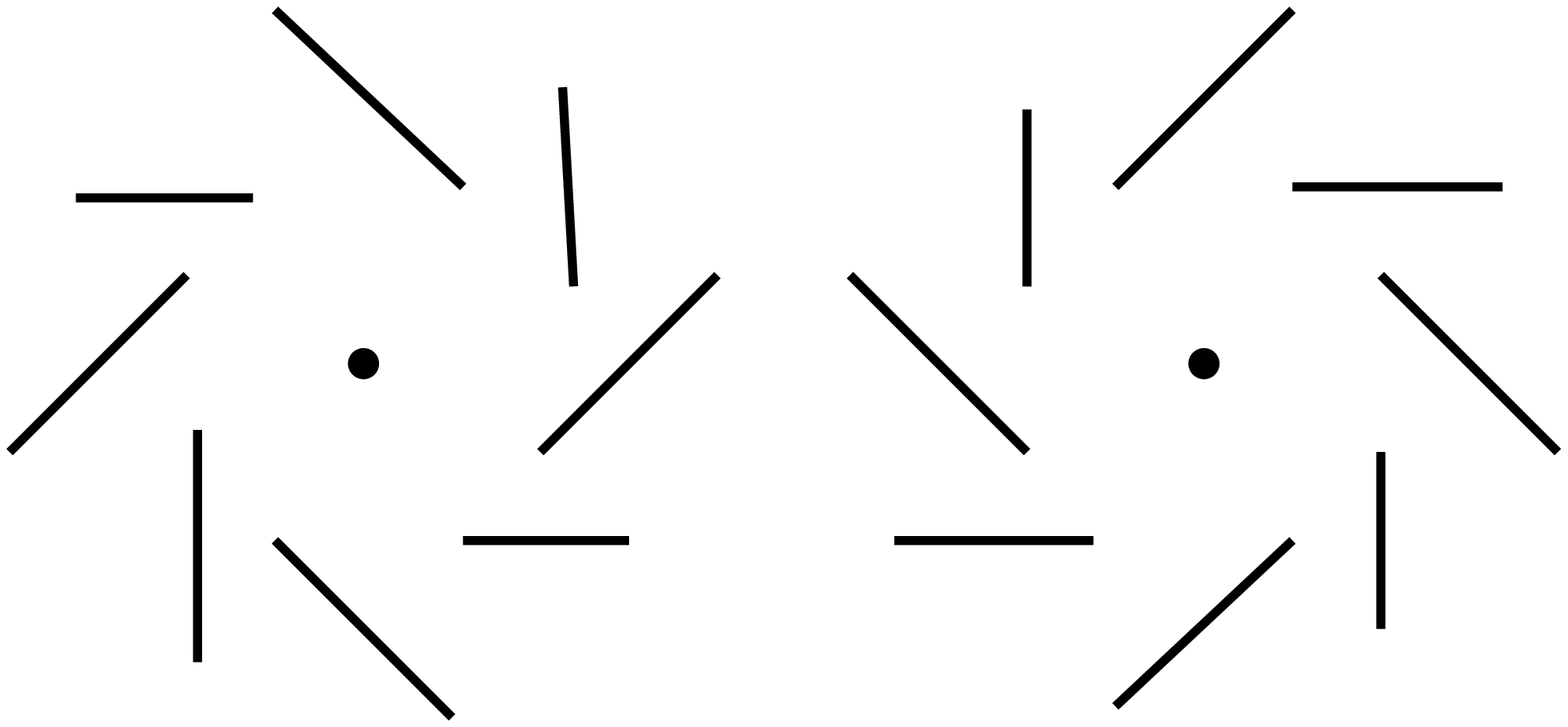,width=4.5in}}
\bigskip
\caption{Polarization pattern with a curl.}
\label{fig:te}
\end{figure}

This can be quantified with a harmonic decomposition of the
polarization field.  The linear polarization of the CMB in 
a direction $\hatn$ is specified by the Stokes parameters
$Q(\hatn)$ and $U(\hatn)$, which are components of a
polarization tensor,
\begin{equation}
  {\cal P}_{ab}(\hatn)={1 \over 2} \left( \begin{array}{cc}
   \vphantom{1\over 2}Q(\hatn) & -U(\hatn) \sin\theta \\
   -U(\hatn) \sin\theta & -Q(\hatn) \sin^2\theta \\
   \end{array} \right),
\label{eq:whatPis}
\end{equation}
which can be thought of as a headless vector.  This polarization 
tensor field can be decomposed into a curl and curl-free part in 
the same way as a vector field can be written in terms of the
gradient of a scalar field plus the curl of some other vector
field; Figs. \ref{fig:sc} and \ref{fig:te} show examples of
gradient and curl polarization patterns, respectively.
Just as the temperature map can be expanded in terms of
spherical harmonics, the polarization tensor can be expanded 
\cite{KamKosSte97a,KamKosSte97b,SelZal97,ZalSel97} (for a
review, see, e.g., Refs.~\cite{cabella,pritchard})
\begin{equation}
      {{\cal P}_{ab}(\hatn)\over T_0} =
      \sum_{lm} \left[ a_{(lm)}^{{\rm G}}Y_{(lm)ab}^{{\rm
      G}}(\hatn) +a_{(lm)}^{{\rm C}}Y_{(lm)ab}^{{\rm C}}(\hatn)
      \right],
\label{eq:Pexpansion}
\end{equation}
in terms of tensor spherical harmonics, $Y_{(lm)ab}^{\rm G}$
and $Y_{(lm)ab}^{\rm C}$, which form a complete orthonormal
basis for the gradient (G) and curl (C) components of the polarization
field (also referred to as ``E'' and ``B'' modes).

The two-point statistics of the combined
temperature/polarization (T/P) map are specified completely by
the six power spectra $C_\ell^{{\rm X}{\rm X}'}=
\VEV{a_{lm}^{\rm X} a_{lm}^{\rm X'}}$,
for ${\rm X},{\rm X}' = \{{\rm T,G,C}\}$ (for temperature,
gradient, and curl, respectively).  Parity invariance
demands that $C_\ell^{\rm TC}=C_\ell^{\rm GC}=0$.
Therefore, the statistics of the CMB
temperature-polarization map are completely specified by the
four sets of moments: $C_\ell^{\rm TT}$, $C_\ell^{\rm TG}$, $C_\ell^{\rm
GG}$, and $C_\ell^{\rm CC}$.

Both density perturbations and gravitational waves will produce
a gradient component in the polarization.  However, only
gravitational waves will produce a curl component
\cite{KamKosSte97a,SelZal97}.  Heuristically,
since density perturbations produce scalar perturbations to the
spacetime metric, they can have no handedness and can thus
produce no curl.  On the other hand, gravitational waves are
propagating disturbances in the gravitational field analogous to 
electromagnetic waves.  A gravitational wave can have right or
left circular polarization, just like an electromagnetic wave.
Gravitational waves can thus carry a handedness, so it
is reasonable that they can produce a polarization pattern with
a handedness, and in fact, they do.  The curl component of the
CMB polarization thus provides a unique signature of the
gravitational-wave background.

Will we ever be able to detect the signature of gravitational 
radiation imprinted on the CMB?  This depends ultimately on the 
height $V$ of the inflaton potential.  Roughly speaking, the raw 
instrumental sensitivity necessary to detect the curl component of the 
polarization from gravitational waves is
\cite{KamKos98,JafKamWan00},
\begin{equation}
    s \lesssim (V^{1/4}/10^{15} \, {\rm GeV} )^{-2}\,  t_{\rm yr}^{1/2} \, 
    \mu{\rm K}~\sqrt{\rm sec},
\end{equation}
where $s$ is the noise-equivalent temperature (NET), which provides a 
measure of the instantaneous sensitivity of the experiment, and
$t_{\rm yr}$ is the duration of the experiment in years. A significant
probe of the GUT parameter space, $V^{1/4} \sim10^{15-16}$ GeV,
will thus require an effective NET approaching 1 $\mu{\rm
K}~\sqrt{\rm sec}$.

\subsection{Slow-roll parameters and gravitational waves}

Once the inflationary potential $V(\phi)$ is specified, the {\it
slow-roll parameters} are defined as
\begin{equation}
     \epsilon= \frac{m_{\rm Pl}^2}{16\pi} \left(
     \frac{V'}{V}\right)^2,
\end{equation}
\begin{equation}
     \eta = \frac{m_{\rm Pl}^2}{8\pi} \frac{V''}{V},
\end{equation}
where the prime denotes derivative with respect to $\phi$.
Slow-roll inflation generally requires $\epsilon,\eta \ll 1$.
In slow-roll inflation, the scalar spectral index (the spectral
index for primordial density perturbations) is $n_s=1-6\epsilon+
2\eta$, and the density-perturbation amplitude determines
$(V/\epsilon)^{1/4}=6.6\times 10^{16}$ GeV.  Thus, $V$, and
therefore the gravitational-wave amplitude, increases with
$\epsilon$.  The commonly used tensor-to-scalar ratio $r=T/S$
(the ratio of the tensor to scalar contributions to the CMB
quadrupole, where tensor here is another term for gravitational
waves) is $r\sim14 \epsilon$.

There have been new developments in the measurement of
inflationary observables with intriguing implications for the
gravitational-wave background.  When combined with other CMB
experiments and large-scale structure, the BOOMERanG 2003 data
suggested $n_s=0.95\pm0.02$ \cite{MacTavish}.  Now, the WMAP
three-year data, when marginalized over a six-dimensional
parameter space, suggest $n_s=0.95 \pm 0.015$, a $3\sigma$
departure from unity \cite{Spergel06}.  For a generic potential,
one expects $\epsilon \sim \eta$.  If so, and if $n_s=0.95$,
then $\epsilon\sim0.01$, and if so, then $V^{1/4}\sim
2\times10^{16}$ GeV and $r\sim0.1$---i.e., the
amplitude of the gravitational-wave background is comparable to
the ``optimistic'' estimates that are usually shown in
experimental-CMB proposals!  In other words, the
gravitational-wave background should be within reach of
next-generation experiments.  Of course, $\epsilon\sim \eta$ is
not guaranteed, and it is in fact possible to construct an inflaton
potential that has $\eta\sim0.01$ and $\epsilon \ll 0.01$.  If
so, then the gravitational-wave background will be small, even
if $n_s=0.95$.  Still, it is perhaps not quite as easy to
construct a model with $\epsilon \ll \eta$ as one might think.
This would require $(V')^2 \ll V''$, a constraint that can be
satisfied only over a narrow range of $\phi$.  As
a specific example, consider the Higgs potential $V(\phi) =
(\phi^2-\mu^2)^2$.  For values of $\phi$ very close to $\phi=0$,
it is indeed true that $\epsilon \ll \eta$.  However, CMB scales
exit the horizon roughly 60 $e$-folds before the end of
inflation.  This constraint demands, for this potential, that
$\phi$ not be too close to the origin, and quantitatively, that
$\epsilon\sim\eta$ leading to a fairly large gravitational-wave
background [as illustrated in Fig.~\ref{fig:inflationbbo}(c)
below].  The bottom line is that although $n_s<1$ does not
``guarantee'' a gravitational-wave background of detectable
amplitude, detection of the gravitational-wave background is
more promising than if $n_s$ had turned out to be consistent with
unity with small error bars.

\vfill\eject
\subsection{Cosmic shear and the CMB}

Although density perturbations produce, in linear theory, no
curl, they can induce a curl component through cosmic shear (CS),
gravitational lensing by density perturbations along the line of
sight \cite{ZalSel98}.  This additional source of curl must be
understood if the CMB polarization is to be used to detect an
inflationary gravitational-wave (IGW) background.  The CS-induced
curl thus introduces a noise from which IGWs must be distinguished.
If the IGW amplitude (or $E_{\rm infl}$) is sufficiently large,
the CS-induced curl will be no problem.  However, as $E_{\rm
infl}$ is reduced, the IGW signal becomes smaller and will at
some point get lost in the CS-induced noise.  If it is not
corrected for, this confusion leads to a minimum detectable IGW
amplitude \cite{LewChaTur02,KesCooKam02,KnoSon02}.

In addition to producing a curl component, CS also introduces
distinct higher-order correlations in the CMB temperature
pattern \cite{SelZal99}.  Roughly speaking, lensing
can stretch the image of the
CMB on a small patch of sky and thus lead to something akin to
anisotropic correlations on that patch of sky, even though the
CMB pattern at the surface of last scatter had isotropic
correlations.  By mapping these effects, the CS can be
mapped as a function of position on the sky \cite{SelZal99}.
The observed CMB polarization can then be corrected for these
lensing deflections to reconstruct the intrinsic CMB
polarization at the surface of last scatter (in which the only
curl component would be that due to IGWs).

Refs.~\cite{KesCooKam02,KnoSon02} show that
if the gravitational-wave background is large enough to be
accessible with the Planck satellite, then the cosmic-shear
contribution to the curl component will not get in the way.
However, to go beyond Planck, the cosmic-shear distortion to the
CMB curl will need to be subtracted by mapping the cosmic-shear
deflection with higher-order temperature-polarization
correlations.  According to the analyses of
Refs.~\cite{KesCooKam02,KnoSon02}, which used quadratic
estimators for the cosmic shear, there will be an irreducible
cosmic-shear-induced curl, even with higher-order correlations,
if the energy scale is $E_{\rm infl}\lesssim 2\times10^{15}$
GeV.  However, maximum-likelihood techniques \cite{hirataseljak}
have been developed for cosmic-shear reconstruction that allow a
reduction in the CS-induced curl by close to two orders of
magnitude below that achievable with quadratic estimators.
Either way, the
cosmic-shear distortions to the CMB will be of interest in their
own right, as they probe the distribution of dark matter
throughout the Universe as well as the growth of density
perturbations at early times.  These goals will be important for
determining the matter power spectrum and thus for testing
inflation and constraining the inflaton potential.

\subsection{CMB and Primordial Gaussianity}

Another prediction of inflation is that the distribution of mass 
in the primordial Universe should be a realization of a Gaussian 
random process.  This means that the distribution of temperature 
perturbations in the CMB should be Gaussian and it moreover
implies a precise relation between all of the higher-order
temperature correlation functions and the two-point correlation
function.  These relations can be tested with future precise CMB
temperature and polarization maps \cite{Veretal00}.  See
Refs.~\cite{cozumel,threept} for reviews.

\subsection{Other implications of CMB results}

Although our focus has been elsewhere, the richness of the
acoustic-peak structure---the locations and heights of the peaks
as well as the troughs---allows the measurements to be used to
simultaneously constrain a number of classical and inflationary
cosmological parameters \cite{Junetal96b,Lanetal00}, in addition to the
total density (determined by the location of the first peak).
CMB maps have now provided an independent and precise new
constraint to the baryon density (verifying the predictions of
big-bang nucleosynthesis \cite{bbn}), robust evidence for
the existence of nonbaryonic dark matter, and an independent
avenue---that confirms supernova evidence \cite{Peretal99}---for
inferring the existence of a cosmological constant.
The CMB results (sometimes combined with large-scale-structure
data) have resulted in a huge number of other new results and
constraints.  One example is the redshift
$z\sim10$ for the formation of the first stars
\cite{WMAP3polarization}.  As three other examples, I mention
precise
constraints to neutrino masses and degrees of freedom (see,
e.g., Refs.~\cite{Pierpaoli,Lesgourgues}), a new constraint to the
amplitude of a primordial gravitational-wave background that
applies to a broad, hitherto unexplored, range of
gravitational-wave frequencies \cite{SmiPieKam06}, and new
constraints to the mass-lifetime-abundance parameter space for
decaying dark-matter particles \cite{decaying}.  In the next
few years, the Planck satellite \cite{Planck} will refine all of
these measurements and constraints to even greater levels of
precision.

\subsection{Direct detection of the gravitational-wave
background?}

\begin{figure}[htbp]
\centerline{\psfig{file=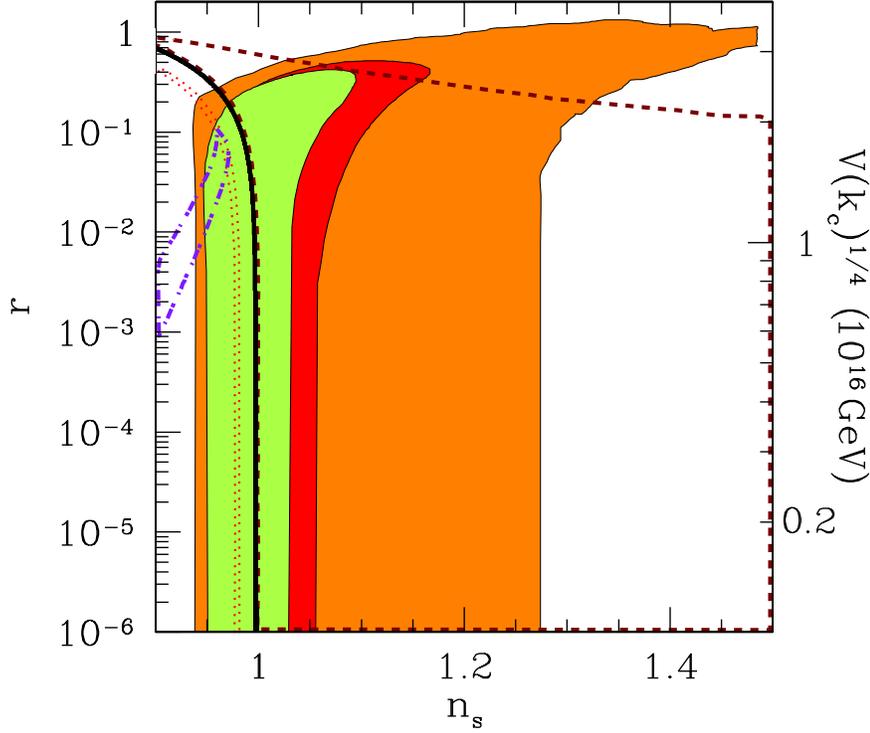,width=4.5in}}
\bigskip
\caption{Regions in the $n_s$--$r$ parameter space consistent
     with the CMB-only (medium gray) \cite{Peiris03}, CMB
     plus galaxy surveys (dark gray), and CMB plus galaxy
     surveys plus Lyman-alpha-forest constraints (light
     gray) \cite{Seljak04}.  Here, $r$ is the tensor-to-scalar
     ratio, and $n_s$ is the scalar spectral index at CMB
     scales.  Plotted on top of these regions are the parameter
     spaces occupied by the four models of inflation we consider:
     power-law (solid line), chaotic (dotted),
     symmetry-breaking (dashed-dot), and hybrid
     (short-dashed).  The parameter space for power-law
     inflation occupies the solid black curve; the parameter
     spaces for the other models occupy the interior of the
     delimited regions.  The right axis shows the energy
     scale $[V(k_c)]^{1/4}$ of inflation. From
     Ref.~\protect\cite{SmiKamCoo05}.  (Note that this Figure
     has now been superseded by Fig.~14 in
     Ref.~\protect\cite{Spergel06}, which restricts further the
     parameter space, favoring a smaller value of $n_s$.  We
     include this older parameter-space plot, as it corresponds
     with the regions shown below in
     Fig.~\protect\ref{fig:inflationbbo}, from the analysis in
     Ref.~\protect\cite{SmiKamCoo05}.)}
\label{fig:rns}
\end{figure}

\begin{figure}[htbp]
\centerline{\psfig{file=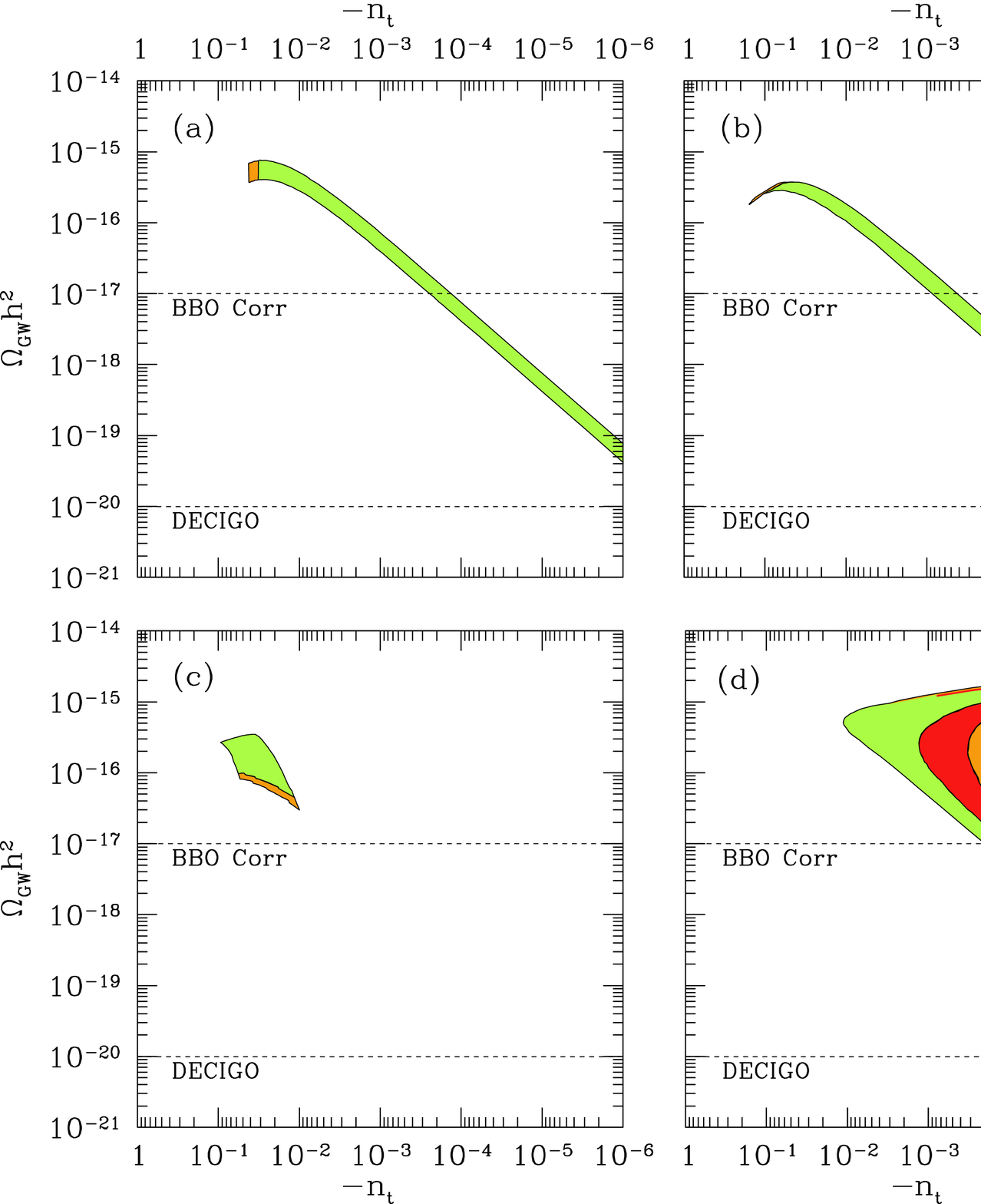,width=4.5in}}
\bigskip
\caption{Regions in the $\Omega_{\rm GW} h^2$--$n_t$ parameter
     space for (a) power-law, (b) chaotic, (c)
     symmetry-breaking, and (d) hybrid inflation.  The shaded
     regions map out the corresponding regions in
     Fig.~\protect\ref{fig:rns}.  Here, the gravitational-wave
     density $\Omega_{\rm GW} h^2$ and spectral index $n_t$ are
     both evaluated at DECIGO/BBO scales.  Also shown are the
     sensitivity goals of BBO and DECIGO.  From
     Ref.~\protect\cite{SmiKamCoo05}.}
\label{fig:inflationbbo}
\end{figure}

If the energy scale of inflation is
high and the IGW spectrum close to scale-invariant, then there is
some prospect for detecting primordial gravitational waves
directly in gravitational-wave observatories (rather than indirectly
through their effect on the CMB), a possibility that has been
considered in Refs.~\cite{Gwdd}.  
Fig.~\ref{fig:spectrum} shows forecasts for sensitivities for
the Big-Bang Observer (BBO) \cite{bbo} and DECIGO (Deci-hertz
Interferometer Gravitational Wave Observatory) \cite{decigo},
two future (i.e., after LISA---Laser Interferometric Space
Antenna---a space-based gravitational-wave detector being
considered now by NASA and ESA) space-based gravitational-wave
detectors that are now under study.  These are families of LISA-like
detectors deployed in the solar system, with ``BBO Corr''
designating a more ambitious configuration in which signals from
various detector arrays are correlated against one another.
DECIGO is an even more ambitious concept.
Ref. \cite{SmiKamCoo05}
considered several classes of inflationary potentials with
parameters chosen to fit CMB constraints, shown in
Fig.~\ref{fig:rns}, to the tensor-to-scalar ratio $r$ (or
equivalently, IGW amplitude) and scalar
spectral index $n_s$.  The shaded regions show consistency
of the parameters with assorted measurements, while the regions
delineated by the lines indicate those regions of parameter
space predicted by various classes of inflationary models.  The
names ``chaotic,'' ``hybrid,'' ``power-law,'' and
``symmetry-breaking'' simply refer to different functional forms
for the inflaton potential; see Ref.~\cite{SmiKamCoo05} for
details.  The predicted gravitational-wave amplitudes
for these four classes of inflationary models
are then shown in Fig.~\ref{fig:inflationbbo}
We see that inflationary models consistent with current data may indeed 
be detectable directly, but detectability depends on the
inflationary model.  It is also difficult to find inflationary
gravitational-wave backgrounds that would be detectable
directly, but not with CMB polarization.  Given the huge
difference in distance scales, detection of the
gravitational-wave background both in the CMB and directly would
provide a powerful lever arm for constraining the inflaton
potential.

\subsection{The CMB polarization: additional remarks}

We have concentrated on CMB polarization as a probe of the
inflationary gravitational-wave background.  However, maps of
the CMB polarization will address a plethora of cosmological
questions.  The small-angle temperature fluctuation is in fact
due to peculiar velocities as well as density perturbations at
the surface of last scatter, while the small-angle polarization
is due only to the peculiar velocity \cite{ZalHar95}.  Thus,
only with a polarization map can primordial perturbations be
reconstructed unambiguously.  The polarization can further constrain the
ionization history of the Universe \cite{Zal97}, help determine the
nature of primordial perturbations \cite{Kos98,SpeZal97}, detect 
primordial magnetic fields \cite{KosLoe96,HarHayZal96,ScaFer97}, 
map the distribution of mass at lower redshifts \cite{ZalSel98},
and perhaps probe cosmological parity violation
\cite{LueWanKam98,Lep98,xuelei06}.

\section{Dark Matter}

Cosmologists have long noted---even well before the recent CMB
results, the discrepancy between the baryon density
$\Omega_b \simeq 0.05$ inferred from BBN and the
nonrelativistic-matter density inferred from cluster masses,
dynamical measurements, and large-scale structure, and the
discrepancy between the baryon and total-matter densities in
galaxy clusters (see, e.g., Ref.~\cite{jkg} for a review of
these pre-CMB arguments).  Today, though, we can simply point to
the exquisite CMB results that suggest a nonbaryonic density
$\Omega_{\rm cdm} h^2 =0.105^{+0.007}_{-0.013}$
\cite{Lanetal00,Spergel06}.  

If neutrinos had a mass $\sim5$ eV, then their density would be
comparable to the dark-matter density.  However, neutrino masses
are now known, from laboratory experiments as well as
large-scale-structure data to be $\lesssim$eV (see, e.g.,
Ref.~\cite{Lesgourgues}); even if neutrinos did have the
right mass, it is difficult to see, essentially from the Pauli
principle \cite{gunn} how they could be the dark matter. It appears
likely then, that some exotic new candidate is required.

For the past two decades, the two leading candidates from
particle theory have been weakly-interacting massive particles
(WIMPs), such as the lightest superpartner (LSP) in
supersymmetric extensions of the standard model
\cite{jkg,bergstrom,hooper}, and axions \cite{axion}.  

\vfill\eject
\subsection{Weakly-interacting Massive Particles}

Suppose that in addition to the known particles of the
standard model, there exists a new stable weakly-interacting
massive particle (WIMP), $\chi$.  At sufficiently early times
after the big bang, when the temperatures are
greater than the mass of the particle, $T\gg m_\chi$, the
equilibrium number density of such particles is $n_\chi \propto
T^3$, but for lower temperatures, $T\ll m_\chi$, the equilibrium
abundance is exponentially suppressed, $n_\chi \propto
e^{-m_\chi/T}$.  If the expansion of the Universe were slow enough
that  thermal equilibrium were always maintained, the number of
WIMPs today would be infinitesimal.  However, the Universe is
not static, so equilibrium thermodynamics is not the entire story.

%
\begin{figure}[htbp]
\centerline{\psfig{file=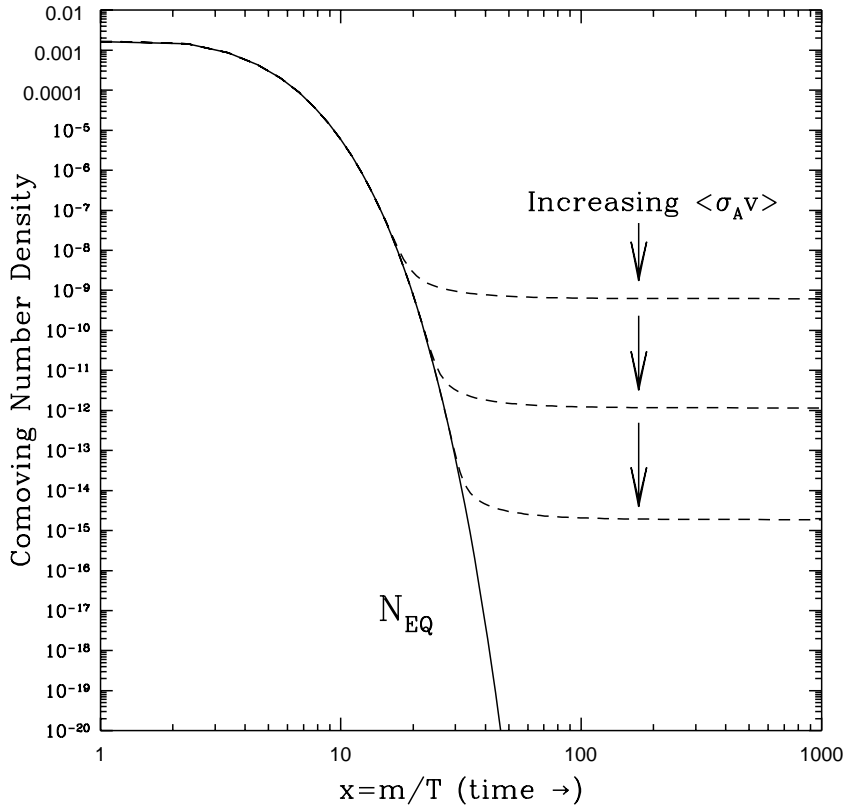,width=4.5in}}
\caption{Comoving number density of WIMPs in the early
        Universe.  The dashed curves are the actual abundances
        for different annihilation cross sections,
        and the solid curve is the equilibrium abundance.  From
        Ref.~\protect\cite{jkg}.}
\label{YYY}
\end{figure}

At high temperatures ($T\gg m_\chi$), $\chi$'s are abundant and
rapidly converting to lighter particles and {\it vice versa}
($\chi\bar\chi\leftrightarrow l\bar l$, where $l\bar l$ are quark-antiquark and
lepton-antilepton pairs, and if $m_\chi$ is greater than the mass of the
gauge and/or Higgs bosons, $l\bar l$ could be gauge- and/or Higgs-boson
pairs as well).  Shortly after $T$ drops below $m_\chi$, the number
density of $\chi$'s drops exponentially, and the rate
$\Gamma=\VEV{\sigma v} n_\chi$ for annihilation of
WIMPs---where $\VEV{\sigma v}$ is the
thermally averaged total cross section $\sigma$ for annihilation
of $\chi\bar\chi$ into lighter particles times relative velocity
$v$---drops below the expansion rate, $\Gamma\la H$.  At this
point, the $\chi$'s cease to annihilate efficiently, they fall
out of equilibrium, and a relic cosmological abundance remains.
The equilibrium (solid line) and actual (dashed
line) abundances of WIMPs per comoving volume are plotted in
Fig.~\ref{YYY} as a function of $x\equiv m_\chi/T$ (which
increases with increasing time).  As the annihilation cross
section is increased, the WIMPs stay in equilibrium longer, so we
are left with a smaller relic abundance when they do finally
freeze out.
An approximate solution to the Boltzmann equation yields the
cosmological WIMP abundance $\Omega_\chi$ (in units of the
critical density $\rho_c$),
\begin{equation}
     \Omega_\chi h^2 ={m_\chi n_\chi \over \rho_c}\simeq
     0.1 \left({3\times 10^{-26}\,{\rm cm}^3 \, {\rm sec}^{-1} \over
     \langle \sigma_A v\rangle }\right).
\label{eq:abundance}
\end{equation}
The result is to a first approximation independent of the WIMP
mass and is fixed primarily by the annihilation cross section.

The WIMP velocities at freeze-out are typically some appreciable
fraction of the speed of light.  Therefore, from
Eq.~(\ref{eq:abundance}), the WIMP will have a cosmological
abundance $\Omega_\chi h^2\sim0.1$ today if the annihilation cross section
is roughly $3\times 10^{-26}\,{\rm cm}^3 \, {\rm sec}^{-1}$, or
in particle-physics units (obtained using $\hbar
c=2\times10^{-14}$~GeV-fm), $10^{-8}$ GeV$^{-2}$.  Curiously,
this is the order of magnitude one would expect from a typical
electroweak cross section,
\begin{equation}
     \sigma_{\rm weak} \simeq {\alpha^2 \over m_{\rm weak}^2},
\end{equation}
where $\alpha \simeq {\cal O}(0.01)$ is the fine-structure
constant, and $m_{\rm weak} \simeq
{\cal O}(100\, {\rm GeV})$.  The numerical constant in
Eq.~(\ref{eq:abundance}) needed to provide $\Omega_\chi h^2\sim0.1$
comes essentially from the expansion rate (which determines the
critical density).  But why should the expansion rate have
anything to do with the electroweak scale?  This remarkable
coincidence suggests that if a
new, as yet undiscovered, stable massive particle with electroweak
interactions exists, then it should have a relic density
suitable to account for the dark matter.  This has been the
argument driving the massive experimental effort to detect WIMPs.

The first WIMPs considered were massive Dirac neutrinos
(particles which have antiparticles) or Majorana
neutrinos (particles that are their own antiparticles) with
masses in the range of a few GeV to a few TeV.
(Due to the Yukawa coupling which gives a neutrino its mass,
neutrino interactions become strong above a few TeV, and the neutrino no
longer remains a suitable WIMP candidate \cite{unitarity}.)  The
Large Electron-Positron (LEP) collider ruled out
neutrino masses below half the $Z^0$ mass.  Furthermore, heavier
Dirac neutrinos have been ruled out as the primary component of
the Galactic halo by direct-detection experiments (described
below) \cite{heidelberg}, and heavier Majorana neutrinos have
been ruled out by indirect-detection experiments
\cite{kamiokande,imb,baksan,macronew,superK,AMANDA} (also
described below) over much of their mass range.  Therefore,
Dirac neutrinos cannot comprise the halo dark matter
\cite{griestsilk}; Majorana neutrinos can, but only over a small
range of fairly large masses.

A much more promising WIMP candidate comes from
electroweak-scale supersymmetry (SUSY)
\cite{jkg,bergstrom,hooper,haberkane}.  SUSY was
hypothesized in particle physics to cure the naturalness problem
with fundamental Higgs bosons at the electroweak scale; in the
GUT theory, the parameter that controls the Higgs-boson mass
must be extremely small, but it may be closer to unity (and
thus, in the particle-theory parlance, more ``natural'') in
supersymmetric theories.
Unification of the strong and electroweak coupling constants at
the GUT scale seems to be
improved with SUSY, and SUSY seems to be an essential ingredient
in theories that unify gravity with the other three fundamental
forces.

The existence of a new symmetry,
$R$-parity, in SUSY theories guarantees that the lightest
supersymmetric particle (LSP) is stable.
In the minimal supersymmetric extension of the
standard model (MSSM), the LSP is usually the neutralino, a linear
combination of the supersymmetric partners of the photon, $Z^0$,
and Higgs bosons.  Another possibility is the sneutrino, the
supersymmetric partner of the neutrino, but
these particles interact like neutrinos and have been ruled out
over most of the available mass range \cite{sneutrino}.  
Given a SUSY model, the cross section for
neutralino annihilation to lighter particles, and thus the relic
density, can be calculated.  The
mass scale of supersymmetry must be of order the weak scale to
cure the naturalness problem, and the neutralino will have only
electroweak interactions.  Therefore, it is to be expected that
the cosmological neutralino density is of order the dark-matter
density, and this is borne out by detailed calculations
in a very broad class of supersymmetric extensions of
the standard model \cite{ellishag}.

\subsection{Direct Detection of WIMPs}

SUSY particles are now among the
primary targets for the Large Hadron Collider (LHC), which
should begin science operations by the end of 2008.  However, one can
also try to detect neutralinos in
the Galactic halo.  In order to account for the dynamics of the
Milky Way, the {\it local} dark-matter density must be $\rho_0
\simeq 0.4\, {\rm GeV}/{\rm cm}^3$, and whatever particles or
objects make up the dark-matter halo must be moving with a
velocity dispersion of 270 km/sec.

Perhaps the most promising technique to detect WIMPs is
detection of the ${\cal O}(30\, {\rm keV})$ nuclear recoil
produced by elastic scattering of neutralinos from nuclei in
low-background
detectors \cite{witten,kim,labdetectors}.  A particle with mass
$m_\chi\sim100$ GeV and electroweak-scale interactions
will have a cross section for elastic scattering from a nucleus
which is $\sigma \sim 10^{-38}\,{\rm cm}^2$.  If the local halo
density is $\rho_0\simeq0.4$ GeV~cm$^{-3}$, and the particles
move with velocities $v\sim 300$ km~sec$^{-1}$, then the rate
for elastic scattering of these particles from, e.g., germanium,
which has a mass $m_N \sim70$ GeV, will be $R \sim \rho_0
\sigma v / m_\chi/m_N \sim1$ event~kg$^{-1}$~yr$^{-1}$.  If a
$100$-GeV WIMP moving at $v/c\sim10^{-3}$ elastically scatters
with a nucleus of similar mass, it will impart a recoil energy
up to 100 keV to the nucleus.  Therefore, if we have 1 kg of
germanium, we expect to see roughly one nucleus per year
spontaneously recoil with an energy of ${\cal O}(30\, {\rm keV})$.

More precise calculations of the detection rate include the
proper neutralino-quark interaction, the QCD and
nuclear physics that turn a neutralino-quark
interaction to a neutralino-nucleus interaction, and a full
integration over the WIMP velocity distribution.  Even if all of these
physical effects are included properly, there is still some
uncertainty in the predicted event rates that arises from
current limitations in our understanding of, e.g., squark,
slepton, chargino, and neutralino masses and mixings.  New
contributions to the neutralino-nucleus cross section are still
being found.  For example, Ref.~\cite{prezeau} found that
there may be a hitherto neglected coupling of the neutralino to
the virtual pions that hold nuclei together.  Rather than
make a single precise prediction, theorists thus generally survey the
available SUSY parameter space.  Doing so, one finds event rates between
$10^{-4}$ to 10 events~kg$^{-1}$~day$^{-1}$ \cite{jkg}, as shown
in Fig.~55 of Ref. \cite{jkg}, although there may be models
with rates that are a bit higher or lower.

\subsection{Energetic Neutrinos from WIMP annihilation}

Energetic neutrinos from WIMP annihilation in the Sun and/or
Earth provide an alternative avenue for indirect detection of
WIMPs \cite{SOS}.
If, upon passing through the Sun, a WIMP scatters elastically from a
nucleus therein to a velocity less than the escape velocity, it
will be gravitationally bound to the Sun.  This leads to a
significant enhancement in the density of WIMPs in the center of
the Sun---or by a similar mechanism, the Earth.  These WIMPs
will annihilate to, e.g., $c$, $b$, and/or $t$ quarks, and/or gauge and
Higgs bosons.  Among the decay products of these particles
will be energetic muon neutrinos that can escape from the
center of the Sun and/or Earth and be detected in neutrino
telescopes such as the Irvine-Michigan-Brookhaven (IMB)
\cite{imb}, Baksan \cite{baksan}, Kamiokande
\cite{kamiokande,superK}, or MACRO \cite{macronew}  (underground
neutrino observatories), or AMANDA \cite{AMANDA} or IceCube
(neutrino observatories built in deep Antarctic ice).
The energies of the neutrino-induced muons will be typically 1/3 to 1/2 the
neutralino mass (e.g., 10s to 100s of GeV), so they will be much
more energetic than ordinary solar neutrinos (and therefore
cannot be confused with them) \cite{JunKam95}.  The signature of such a
neutrino would be the Cerenkov radiation emitted by an upward
muon produced by a charged-current interaction between the
neutrino and a nucleus in the material below the detector.

The annihilation rate of these WIMPs equals the rate for
capture of these particles in the Sun \cite{pressspergel}.  The flux of
neutrinos at the Earth depends also on the Earth-Sun distance,
WIMP-annihilation branching ratios, and the decay branching
ratios of the annihilation products.  The flux of upward muons
depends on the flux of neutrinos and the cross section for
production of muons, which depends on the square of the neutrino
energy.  

As in the case of direct detection, the precise
prediction involves numerous factors from particle and nuclear
physics and astrophysics, and on the SUSY parameters.
When all these factors are taken into account, predictions for
the fluxes of such muons in SUSY models
seem to fall for the most part between $10^{-6}$ and 1
event~m$^{-2}$~yr$^{-1}$ \cite{jkg}, as shown in
Fig.~57 of Ref. \cite{jkg}, although the numbers may be a bit
higher or lower in some models.  Presently, IMB, Kamiokande
Baksan, and MACRO constrain the flux of energetic neutrinos from
the Sun to be $\lesssim0.02$~m$^{-2}$~yr$^{-1}$
\cite{kamiokande,imb,baksan,macronew}.  Larger and more
sensitive detectors such as super-Kamiokande \cite{superK} and
AMANDA \cite{AMANDA} are now operating, and others are being
constructed \cite{IceCube}.

\subsection{Recent Results}

The experimental effort to detect WIMPs began nearly twenty
years ago, and the theoretically favored regions of the SUSY
parameter space are now beginning to be probed.
An earlier claimed detection by the DAMA collaboration
\cite{dama} has been shown to be in conflict with null
searches from the EDELWEISS \cite{edelweiss}, ZEPLIN \cite{zeplin},
and Cryogenic Dark Matter Search (CDMS) \cite{cdms06}
experiments, if the WIMP couples to the mass of the nucleus, and
it is conflict with CDMS \cite{cdmsspin} if it couples instead
to nuclear spins.  The putative DAMA signal
also conflicts, under a
fairly broad range of assumptions, with energetic-neutrino
searches \cite{piero,modelindependent,kurylov}.  WIMPs have not
yet been discovered, but only a small region of the parameter
space has yet been probed.  It will take another generation of
experiments to probe the favored parameter space.

\subsection{WIMPs and exotic cosmic rays}

WIMPs might also be detected via observation of exotic
cosmic-ray positrons, antiprotons, and gamma rays produced by
WIMP annihilation in the Galactic halo.  The difficulty
with these techniques is discrimination between WIMP-induced
cosmic rays and those from traditional astrophysical
(``background'') sources. However, WIMPs may
produce distinctive cosmic-ray signatures.  
For example, WIMP annihilation might produce a
cosmic-ray-positron excess at high energies \cite{baltz,positrons}.
There are now several balloon (e.g., BESS,
CAPRICE, HEAT, IMAX, MASS, TS93) and satellite (AMS and PAMELA)
experiments that have recently flown or are about to be flown to search for
cosmic-ray antimatter.  In fact, the HEAT experiment may already
show some evidence for a positron excess at high energies
\cite{heatpositrons}.  

WIMP annihilation will produce an antiproton excess at low
energies \cite{JunKam94}, although Ref. \cite{pieropbar} claims
that more traditional astrophysical sources can mimic such an excess.
They argue that the antiproton background 
at higher energies ($\ga$few GeV) is better understood, and that
a search for an excess of these higher-energy antiprotons would
thus provide a better WIMP signature.  Cosmic-ray antideuterons
have also been considered as a signature of WIMP annihilation
\cite{deuterons}.

Direct WIMP annihilation to two photons can produce a gamma-ray
line, which could not be mimicked by a traditional astrophysical
source, at an energy equal to the WIMP mass.  WIMPs could also
annihilate directly to a photon and a $Z^0$ boson
\cite{berkap,pieroZphoton}, and these photons will be
monoenergetic with an energy that differs from that of the
photons from direct annihilation to two  photons.  Resolution of
both lines and measurement of their
relative strengths would shed light on the composition of the
WIMP.  Ground-based experiments like VERITAS, HESS, STACEE,
CELESTE, or CACTUS or the GLAST satellite will seek this
annihilation radiation.  A recent (null) search was carried out
for WIMP-annihilation lines in EGRET data \cite{anthony}.

It was recently argued \cite{GS} that there may be a very
dense dark-matter spike, with a dark-matter density that scales
with radius $r$ as $\rho(r) \propto r^{-2.25}$ from the Galactic
center, around the black hole at the Galactic center.  If so, it
would give rise  to a huge flux of annihilation radiation.
However, others have questioned whether this spike really arises
\cite{zhao}.

While the Galactic center provides one source for gamma rays
from WIMP annihilation, it has also been argued that other
sources---in particular, the Draco dwarf galaxy---may have a
sufficiently dense dark-matter core to provide an alternative
target for WIMP-induced gamma rays \cite{draco}.  A tentative
excess of $\sim100$-GeV gamma rays from
Draco \cite{cactusexpt}.  was shown \cite{cactus} shown to
require WIMP-annihilation cross sections that are most likely
too high to be explained by supersymmetric models, unless the
central dark-matter halo of Draco has a very steep cusp.

\subsection{Non-minimal WIMPs?}

N-body simulations of structure formation with collisionless
dark matter show dark-matter cusps, density profiles that fall
as $\rho(r)\propto 1/r$ with radius $r$ near the galactic
center \cite{nfw}, while some dwarf-galaxy rotation curves
indicate the existence of a density core in their centers
\cite{moore}.  This has prompted some theorists to consider
self-interacting dark matter \cite{ss}.  If dark-matter
particles elastically scatter from each other in a galactic
halo, then heat can be transported from the halo center to the
outskirts, thereby smoothing the cusp into a core.
In order for this mechanism to work, however, the
elastic-scattering cross section must be $\sigma_{\rm el} \sim
10^{-(24-25)} (m_\chi/{\rm GeV})$ cm$^2$, roughly thirteen
orders of magnitude larger than the cross section expected for
WIMPs, and even further from that for axions.  If the cross
section is stronger, the halo will undergo core collapse
\cite{corecollapse}, and if it is weaker, the heat transport is
not efficient enough to remove the dwarf-galaxy
dark-matter cusp.

The huge discrepancy between the magnitude of the required
scattering cross section and that for WIMPs and axions has made
self-interacting dark matter unappealing to most WIMP and axion
theorists (but see, e.g., Refs. \cite{mcdonald}).  However,
theoretical prejudices aside, self-interacting dark matter now
seems untenable observationally.  If dark matter is collisional,
dark-matter cores should equilibrate and become round.
Non-radial arcs in the gravitational-lensing system MS2137-23
require a non-spherical core and thus rule out the scattering
cross sections required to produce dwarf-galaxy cores
\cite{jordi}.  One possible loophole is that the
scattering cross section is inversely proportional to the
relative velocity of the scattering particles; this would
lengthen the equilibration time in the core of the cluster
MS2137-23.  This possibility has now been ruled out, however, by
x-ray observations of the giant elliptical galaxy NGC 4636 which
shows a very dense dark-matter cusp at very small radii
\cite{loewenstein}.

There are (many!)~other ways that non-minimal WIMPs could make
themselves manifest cosmologically and astrophysically.  As one
example, Ref. \cite{KrisMarc} we considered the effects of
WIMPs that are produced via decay of a charged particle with a
lifetime of 3.5 years.  If a WIMP spends the first 3.5 years of
its existence as a charged particle, then during that time it
couples to the baryon-photon plasma in the early Universe.  If
so, then pressure support from the plasma prevents the
gravitational amplification of density perturbations in the WIMP
fluid.  Thus, the growth of modes that enter the horizon during
the first 3.5 years---i.e., those on sub-Mpc comoving
scales---is suppressed.  This suppression can then explain the
dearth of dwarf galaxies in the Local Group \cite{KamLid}.
Although not generic, this
charged-particle decay can occur in supersymmetric
models \cite{StefanoKrisMarc}, and there are ways, with 21-cm
probes of the high-redshift Universe, that this mechanism
may be distinguished from those \cite{KamLid} where the
suppression is introduced by broken scale invariance during
inflation.

\subsection{Kaluza-Klein modes and other possibilities}

Inspired by the presumed existence of extra spatial dimensions,
it has become quite fashionable among particle theorists in
recent years to consider the possibility that the Universe may
contain large extra dimensions in which the graviton may travel,
but which are inaccessible to standard-model fields.  The array
of models and phenomenology that has been derived from them is
startling.  However, there is a subclass of these theories, {\it
universal extra dimensions} (see Ref.~\cite{hooperprofumo} for a
recent review), in which standard-model fields are
allowed to propagate on a toroidal compact extra dimension,
usually taken to have a size $d\sim{\rm TeV}^{-1}$.  The momenta
in these extra dimensions are quantized in units of $\hbar /
(2\pi d)$ and appear in our
3+1-dimensional space as a mass.  What this
means is that for every standard-model particle, there is
a series of particles, ``Kaluza-Klein'' excitations (named after
Kaluza and Klein, who first studied extra spatial dimensions),
with the same quantum numbers and masses
close to the inverse size of the extra dimension.  The lightest
of these KK modes is stable, due to conservation of momentum in
the extra dimension.  These particles can annihilate with
particles with the opposite quantum numbers and opposite
momenta in the extra dimension, with interaction
strengths characteristic of the electroweak scale, and they may
elastically scatter from ordinary particles, also with
electroweak-strength interactions.  Consequently, the
dark-matter phenomenology of these particles parallels quite
closely that of supersymmetric WIMPs.

Another avenue recently explored is to consider WIMPs in a
model-independent way.  In particular, there are obvious
phenomenological questions one can ask, such as how dark is
``dark''?  I.e., how weak must the coupling of the photon be to
the WIMP?  One way to answer
this question is to postulate that the WIMP has a tiny
electromagnetic charge, a millicharge, and then constrain the
value of the charge as a function of its mass
\cite{millicharge}.  Another possibility is to suppose the
dark-matter particle is neutral, but couples to the photon
through an electric or magnetic dipole \cite{dipole}.

\subsection{Kinetic decoupling of WIMPs and small-scale
structure}

When we speak of freeze-out of WIMPs in the early Universe, we
usually refer to the freezing out of WIMP annihilation and thus
the departure of WIMPs from
{\em chemical} equilibrium.  This, however, does \emph{not}
signal the end of WIMP interactions.  {\it Elastic} scattering of
WIMPs from light standard-model particles in the primordial
plasma keep WIMPs in \emph{kinetic} equilibrium until later
times (lower temperatures)
\cite{Boehm,Chen:2001jz,GreenWIMPy}. The temperature $T_{\rm
kd}$ of {\em kinetic} decoupling sets the
distance scale at which linear density perturbations in the
dark-matter distribution get washed out---the small-scale 
cutoff in the matter power spectrum.  In turn, this small-scale
cutoff sets the mass $M_c \simeq 33.3\left(T_{\rm kd}/10\ {\rm
MeV}\right)^{-3}~M_{\oplus}$ \cite{Loeb:2005pm} (where $M_\oplus$
is the Earth mass) of the
smallest protohalos that form when these very small scales go
nonlinear at a redshift $z \sim 70$.  There may be implications
of this small-scale cutoff for direct \cite{Diemand:2005vz} and
indirect \cite{Ando:2005xg} detection.

Early work assumed that the cross sections for WIMPs to
scatter from light particles (e.g., photons and neutrinos) would
be energy independent, leading to suppression of power out to
fairly large (e.g., galactic) scales.  However, in
supersymmetric models, at least, the relevant elastic-scattering
cross sections drop precipitously with temperature, resulting in
much higher $T_{\rm kd}$ and much smaller suppression
scales \cite{Chen:2001jz}.  This estimate has
been used to derive $T_{\rm kd}$ and infer
that the minimum protohalo mass is $M_c \sim M_{\oplus}$
\cite{GreenWIMPy,Loeb:2005pm,Diemand:2005vz}.  

Ref.~\cite{ourkinetic} calculated the kinetic-decoupling
temperature $T_{\rm kd}$ of supersymmetric and UED dark matter
concluding that $T_{\rm kd}$ may range all the way from tens of
MeV to several GeV implying a range $M_c\sim10^{-6}~
M_{\oplus}$ to $M_c\sim10^{2}~M_{\oplus}$.  

\subsection{Axions}

The other leading dark-matter candidate is the axion
\cite{axion}. The QCD Lagrangian may be written
\begin{equation}
     {\cal L}_{QCD} = {\cal L}_{\rm pert} + \theta {g^2 \over 32
     \pi^2} G \widetilde{G},
\end{equation}
where the first term is the perturbative Lagrangian responsible
for the numerous phenomenological successes of QCD.  However,
the second term (where $G$ is the gluon field-strength tensor
and $\widetilde{G}$ is its dual), which is a consequence of
nonperturbative effects, violates charge-parity ($CP$) symmetry.
From constraints to the
neutron electric-dipole moment, $d_n \la 10^{-25}$ e~cm, it can
be inferred that $\theta \la 10^{-10}$.  But why is $\theta$ so
small?  This is the strong-$CP$ problem.

The axion arises in the Peccei-Quinn (PQ) solution to the strong-$CP$
problem \cite{PQ}.  A global $U(1)_{PQ}$ symmetry is broken at a
scale $f_a$, and $\theta$ becomes a dynamical field with a
flat potential.  At temperatures
below the QCD phase transition, nonperturbative quantum effects
break explicitly the symmetry and produce a non-flat potential
that is minimized at $\theta\rightarrow 0$.
The axion is the pseudo-Nambu-Goldstone boson of this
near-global symmetry, the particle associated with excitations
about the minimum at $\theta=0$.  The axion mass is $m_a \simeq\, {\rm
eV}\,(10^7\, {\rm GeV}/ f_a)$, and its coupling to ordinary
matter is $\propto f_a^{-1}$.

The Peccei-Quinn solution works equally well for
any value of $f_a$.  However, a variety
of astrophysical observations and laboratory experiments
constrain the axion mass to be $m_a\sim10^{-4}$ eV.
Smaller masses would lead to an
unacceptably large cosmological abundance.  Larger masses
are ruled out by a combination of constraints from supernova
1987A, globular clusters, laboratory experiments, and a search
for two-photon decays of relic axions.

Curiously enough, if the axion mass is in the relatively small viable
range, the relic density is $\Omega_a\sim1$, and so the axion may
account for the halo dark matter.  Such axions would be produced
with zero momentum by a misalignment mechanism in the early
Universe and therefore act as cold dark matter.  During the process of
galaxy formation, these axions would fall into the Galactic
potential well and would therefore be present in our halo with a
velocity dispersion near 270 km~sec$^{-1}$.

It has been noted that quantum gravity is generically expected
to violate global symmetries, and unless these Planck-scale
effects can be suppressed by a huge factor, the Peccei-Quinn
mechanism may be invalidated \cite{gravity}.  Of course, we have
at this point no predictive
theory of quantum gravity, and several mechanisms for forbidding
these global-symmetry violating terms have been proposed
\cite{solutions}.  Therefore, discovery of an
axion might provide much needed clues to the nature of
Planck-scale physics.

There is a very weak coupling of an axion to photons through the
triangle anomaly, a coupling mediated by the exchange of virtual
quarks and leptons.  The axion can therefore decay to two
photons, but the lifetime is $\tau_{a\rightarrow \gamma\gamma}
\sim 10^{50}\, {\rm s}\, (m_a / 10^{-5}\, {\rm eV})^{-5}$ which
is huge compared to the lifetime of the Universe and therefore
unobservable.  However, the $a\gamma\gamma$ term in the
Lagrangian is ${\cal L}_{a\gamma\gamma} \propto a {\vec E} \cdot
{\vec B}$ where ${\vec E}$ and ${\vec B}$ are the electric and
magnetic field strengths.  Therefore, if one immerses a resonant
cavity in a strong magnetic field, Galactic axions that pass
through the detector may be converted to fundamental excitations
of the cavity, and these may be observable \cite{sikivie}.  Such
an experiment is currently underway \cite{axionexperiments} and
has already begun to probe part of the cosmologically
interesting parameter space (see the Figure in Ref.~\cite{karlles}), and it
should cover most of the interesting region parameter space in
the next few years. 

Axions, or other light pseudoscalar particles, may show up
astrophysically or experimentally in other ways.  For example,
the PVLAS Collaboration \cite{pvlas} reported the observation of
an anomalously large rotation of the linear polarization of a
laser when passed through a strong magnetic field.  Such a
rotation is expected in quantum electrodynamics, but the
magnitude they reported was in excess of this expectation.  One
possible explanation is a coupling of the pseudoscalar $F \tilde
F$ of electromagnetism to a low-mass axion-like pseudoscalar
field.  The region of the mass-coupling parameter space implied
by this experiment violates limits for axions from astrophysical
constraints, but there may be nonminimal models that can
accommodate those constraints.  Ref. \cite{kris} reviews the
theoretical interpretation and shows how the PVLAS results may
be tested with x-ray re-appearance experiments.

\section{Dark Energy}

In addition to confirming the predictions of big-bang
nucleosynthesis and the existence of dark matter, the
measurement of classical cosmological parameters has resulted in
a startling discovery: roughly 70\% of the energy density
of the Universe is in the form of some mysterious
negative-pressure dark energy \cite{Carroll:2000fy}.  The
original supernova
evidence for an accelerating Universe~\cite{Peretal99}
has now been dramatically bolstered by CMB measurements, which
indicate a vacuum-energy contribution $\Omega_\Lambda\simeq0.7$ to
the critical density.


As momentous as these results are for
cosmology, they may be even more remarkable from the vantage point of
particle physics, as they indicate the existence of new physics
beyond the standard model plus general relativity.  Either
gravity behaves very peculiarly on the very largest scales, and/or
there is some form of negative-pressure ``dark energy'' that
contributes 70\% of the energy density of the Universe.
As shown below, if this dark energy is to accelerate the expansion, its
equation-of-state parameter ${w}\equiv p/\rho$ must be
${w}<-1/3$, where $p$ and $\rho$ are the dark-energy pressure
and energy density, respectively.  The simplest guess for this dark
energy is the spatially uniform time-independent cosmological
constant, for which ${w}=-1$. Another possibility is
quintessence~\cite{Caletal98} or
spintessence \cite{BoyCalKam01}, a cosmic scalar field that is
displaced from the minimum of its potential.  Negative pressure is
achieved when the kinetic energy of the rolling field is less than the
potential energy, so that $-1 \le {w} < -1/3$ is possible.

The dark energy was a complete surprise and remains a
complete mystery to theorists, a stumbling block that, if
confirmed, must be understood before a consistent unified theory can be
formulated.  This dark energy may be a direct remnant of string
theory, and if so, it provides an exciting new window to
physics at the Planck scale.

Although it is the simplest possibility, a cosmological constant
with this value is strange, as quantum gravity would
predict its value to be $10^{120}$ times the observed value, or
perhaps zero in the presence of some symmetry.  
One of the appealing features of dynamical models for dark
energy is that they may be compatible with a true vacuum
energy which is precisely zero, to which the Universe will
ultimately evolve.  

\subsection{Basic considerations}

The first law of thermodynamics (conservation of energy) tells
us that if the Universe is filled with a substance of pressure
$p=w\rho$, where $\rho$ is the energy density and $w$ the
equation-of-state parameter, then the change in the energy
$dE=d(\rho a^3)$ in a comoving volume (where $a$ is the scale
factor) is equal to the work $dW= - p d(a^3)$ done by the
substance.  Some algebraic rearrangement yields $(d\rho/\rho)=
-3(1+w) (da/a)$ from which it follows that the energy density of
the substance scales as $\rho \propto a^{-3(1+w)}$.  
For example, nonrelativistic matter has $w=0$ and $\rho\propto
a^{-3}$, while radiation has $w=1/3$ and $\rho\propto a^{-4}$.
And if $w=-1$, we get a cosmological constant, $\rho
\propto$constant.  Now in order to get cosmic acceleration, we
require superluminal expansion; that is, that the scale factor
$a$ grow more rapidly than $t$.  If the Universe is filled with
a substance with equation of state $p=w\rho$, then the Friedmann
equation is $H \propto (\dot a/a) \propto a^{-3(1+w)}$, from
which it follows that $a\propto t^{-2/3(1+w)}$.  We thus infer
that we must have $w<-1/3$ for cosmic acceleration.

A negative pressure may at first be counterintuitive, but
intuition is rapidly established when we realize that a negative
pressure is nothing but tension---i.e., something that pulls,
like a rubber band, rather than pushes, like the molecules in a
gas.  Still, one may then wonder how it is that something that
pulls can lead to (effectively) repulsive gravity.  The answer
is simple.  In Newtonian mechanics, it is the mass density
$\rho$ that acts as a source for the gravitational potential
$\phi$ through the Poisson equation $\nabla^2\phi = 4 \pi G
\rho$.  In general relativity, it is energy-momentum that sources the
gravitational field.  Thus, in a molecular gas, pressure, which
is due to molecular momenta, can also source the
gravitational field.  Roughly speaking, the Newtonian Poisson
equation gets replaced by $\nabla^2 \phi = 4\pi G (\rho +3p)$.
Thus, if $p<-\rho/3$, gravity becomes repulsive rather than
attractive.

\subsection{Observational probes}

The obvious first step to understand the nature of this dark
energy is to determine whether it is a true cosmological
constant ($w=-1$), or whether its energy density evolves with
time ($w\neq-1$).  This can be 
answered by determining the expansion rate of the Universe as a
function of redshift.  In principle, this can be accomplished
with a variety of cosmological observations (e.g.,
quasar-lensing statistics, cluster abundances and properties,
the Lyman-alpha forest, galaxy and cosmic-shear surveys, etc.).
However, the currently leading contenders in this race are
supernovae, particularly those that can reach beyond
redshifts $z\gtrsim1$.  Here, better systematic-error reduction,
better theoretical understanding of supernovae and
evolution effects, and greater statistics, are all required.
Both ground-based (e.g., the LSST \cite{LSST}) and space-based
(e.g., SNAP/JDEM \cite{SNAP})
supernova searches can be used to determine the expansion
history.  However, for redshifts $z\gtrsim1$, the principal optical
supernova emission (as well as the characteristic silicon
absorption feature) gets shifted to the infrared which is
obscured by the atmosphere.  Thus, a space-based observatory
appears to be desirable to reliably measure the expansion
history in the crucial high-redshift regime.

In recent years, baryon acoustic oscillations have become
increasingly attractive as a possibility for determining the
expansion history.  The acoustic oscillations seen in the CMB
power spectrum are due to oscillations in the photon-baryon
fluid at the surface of last scatter.  The dark matter is
decoupled and does not participate in these oscillations.
However, since baryons contribute a non-negligible fraction of
the nonrelativistic-matter density, oscillations in the
baryon-photon fluid get imprinted as small oscillations in the
matter power spectrum at late times \cite{oscillations}.  Quite
remarkably, these oscillations have now been detected in galaxy
surveys \cite{sdssbao}.  The
physical distance scale at which these oscillations occur is
well understood from linear perturbation theory, and they thus
provide a standard ruler.  The effects of cosmological geometry
can therefore be inferred by comparing their observed angular
size to that expected from their distance.  If this can be done
at a variety of redshifts,
including high redshifts $z\gtrsim1$, then these acoustic
oscillations provide a way to 
measure the expansion history \cite{eisenstein}.  There are now
a number of competing proposals and efforts to carry out galaxy
surveys at high redshifts to make these measurements.

The other two leading candidates for expansion-history probes are
cluster surveys and cosmic-shear (weak
gravitational lensing) surveys, but there are many others that
have been proposed.  For example, the abundance of
proto-clusters, massive overdensities that have yet to virialize
and become x-ray clusters, has been suggested as a dark-energy
probe \cite{nevinone}.  Another suggestion is to measure to
relative ages of cluster ellipticals as a function of redshift
\cite{aviraul}.  

\subsection{Supernova data}

\begin{figure}[htbp]
\centerline{\psfig{file=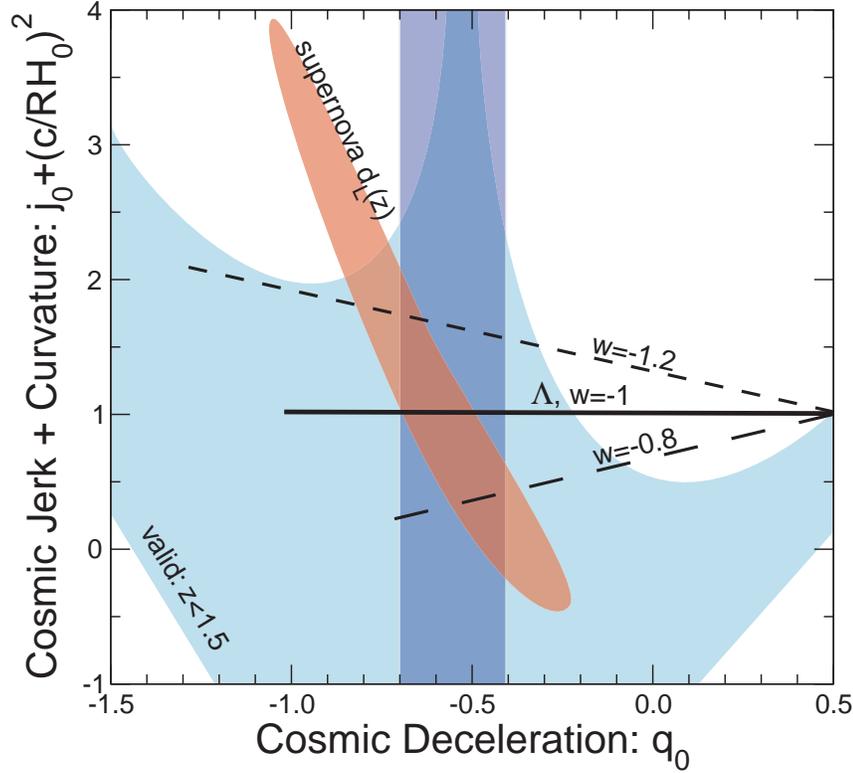,width=4.5in}}
\caption{Current constraints to the $[q_0,\, j_0+(H_0 R)^{-2}]$
     plane, where $R$ is the universal radius of curvature. The dark
     shaded region is the $95\%$ confidence-level constraint from recent
     high-redshift supernova measurements \protect\cite{riess}.  The
     light-shaded region shows the domain of validity of the cubic redshift
     expansion; more precisely, outside these regions, there would be a unit
     magnitude error at $z=1.5$ introduced by the quartic term.    The solid
     curve indicates a family of flat cosmological-constant models with
     decreasing matter density from right to left, terminating at $q_0=-1$ when
     $\Omega_m=0$.  The short-dash curve shows the same for flat models with
     quintessence with $w=-1.2$, and the long-dash curve shows the
     same for $w=-0.8$.  The vertical band shows the range of values for a
     spatially-curved model with $\Omega_m+\Omega_\Lambda=1$
     and matter density spanning the range $0.2 < \Omega_m
     <0.4$.  From Ref.~\protect\cite{robertmarc}.}
\label{fig:jerk}
\end{figure}

The supernova statistics have been building steadily since the
initial 1998 results.  Two years ago, it was announced that
supernova data at high redshift were able to see the transition
between cosmic acceleration and cosmic deceleration expected
at earlier times \cite{riess}.  More precisely, the measurements
of the luminosity-distance--redshift relation (the relation
between the distances inferred by the apparent brightness of
``standard candles,'' sources of fixed luminosity)  had become
sufficiently precise to measure the cosmic jerk $j_0$, the cubic
correction to the expansion law, in addition to the usual
deceleration parameter $q_0$, the quadratic correction.
Ref.~\cite{robertmarc} pointed out that this measurement
provides the first classical (i.e., non-CMB) cosmological probe
of the geometry of the Universe.  The point is that the spatial
curvature in Friedmann-Robertson-Walker (FRW) models does not
enter until the cubic term in
the expressions for the angular-diameter distance (the distance
inferred by the observed angular size of an object of known
physical size) and luminosity distance.  Assuming, then, that
the dark energy is a cosmological constant allows us to use
these results to constrain the curvature scale, as shown in
Fig.~\ref{fig:jerk}.

\subsection{Quintessence}

The simplest paradigm for cosmic acceleration is quintessence.
The idea is somewhat similar to inflation.  In such scenarios,
one postulates a scalar field $\phi(t,\vec x)$ with a
potential-energy density $V(\phi)$, such that the scalar field
is rolling sufficiently slowly down its potential to lead to an
accelerated expansion.  The equation of motion for the
homogeneous component of the field is $\ddot \phi +3 H \dot \phi
+V'(\phi)=0$, where the dot denotes derivative with respect to
time, and $H$ is the expansion rate.  Here, the expansion serves
as a friction term that prevents the scalar field from rolling
directly to its minimum.  The pressure in the field is $p =
(1/2) \dot\phi^2 -V(\phi)$, and the energy density is $\rho
=(1/2)\dot\phi^2+V(\phi)$.  Thus, if the field rolls slowly
enough, then $w<-1/3$ and cosmic acceleration can proceed.

Quintessence models can be designed to provide the correct
energy density today, but the right answer usually has to be put
in by hand.  As with the cosmological constant, the ``why now''
problem---i.e., why does the vacuum energy show up billions of
years after the big bang, rather than much earlier or
later?---is not really answered.  There may be ``tracker models,''
\cite{tracker} though, that go some way toward addressing this
problem.  It turns out that if the quintessence potential is
$V(\phi) \propto e^{-\phi/\phi_0}$, then during matter or
radiation domination, the field rolls down the potential in such
a way that the kinetic-plus-potential energy density scales with
the expansion in the same way as the dominant component, matter
or radiation.  Thus, the scalar-field energy density in such
models is not required to be infinitesimal compared with the
dominant energy component over many decades in scale factor.

Another class of alternatives includes spintessence
\cite{BoyCalKam01}, in which one postulates a complex scalar
field with a $U(1)$ symmetry.  The field is then postulated to
be spinning in the $U(1)$ symmetric potential, and it is the
centrifugal-force barrier (or the conserved global charge),
rather than expansion friction, that prevents the field from
rolling directly to its minimum.  Depending on the form of the
potential $V(|\phi|)$, spintessence can act as dark matter or as
dark energy.  There is, however, generically an instability to
production of Q-balls (balls of spinning scalar field) for
spintessence potentials that produce
cosmic acceleration, and finding workable spintessence models
for acceleration has proved to be difficult.

The astronomical observations aimed at probing dark energy aim,
to a first approximation, to determine the expansion history of
the Universe.  A few may probe the possible effects of
quintessence or other models on the growth of perturbations,
particularly on large scales.  However, might there be other
ways to determine the physics of dark energy?
If the dark energy is quintessence, rather than a cosmological
constant, then there may be observable consequences in the
interactions of elementary particles if they have some coupling
to the quintessence field.  In particular, if the cosmological
constant is time evolving (i.e., is quintessence), then there is
a preferred frame in the Universe.  If elementary particles
couple weakly to the quintessence field, they may exhibit small
apparent violations of Lorentz and/or CPT symmetry (see, e.g.,
Ref.~\cite{Carroll}).  A variety of accelerator and astrophysical
experiments \cite{Carroll,LueWanKam98,Lep98} can be done to
search for such exotic signatures.

Quintessence models are simple and fairly predictive, once the
potential $V(\phi)$ is specified.  Although they must all be
considered toy models, they are handy as working
phenomenological models, or placeholders for a more fundamental
theory.

\vfill\eject
\subsection{Alternative gravity}

Quintessence postulates the existence of some new form of
``dark energy,'' a scalar-field configuration, with negative pressure
that then drives the accelerated expansion in accord with
general relativity.  Another possibility is that there is no new
exotic substance, but that the laws of gravity are modified on
large distance scales.  One simple example is $1/R$ gravity
\cite{1overR}.  The usual Einstein-Hilbert Lagrangian is simply
proportional to the Ricci scalar $R$, which measures the scalar
curvature of space.  When this action is minimized, it leads to
Einstein's equation.  In the absence of matter, the isotropic
homogeneous spacetime that minimizes the action is Minkowski space;
i.e., a spacetime with $R=0$. If, however, we postulate an
additional term, $\mu^4/R$, where $\mu$ is a (very small) mass
scale, in the action, then the isotropic homogeneous spacetime
that minimizes the action is $R=\mu^2$ \cite{1overR}; i.e., de Sitter space.
Thus, an empty Universe has an accelerated expansion, and a
sufficiently low-density Universe, like our own, is headed
toward a de Sitter spacetime.  Unfortunately, though, this model
is phenomenologically untenable \cite{chiba}.  Theories
in which the action is a function $f(R)$ of the Ricci scalar can
be mapped onto scalar-tensor theories.  The additional term in
the action brings to life the scalar degree of freedom in the
metric, leading to a change in the spacetime metric surrounding
a massive object.  Thus, the deflection of light by the Sun is
altered in a way that is (very) inconsistent with current
limits.

An alternative approach comes from large extra dimensions.  In
DGP (for Dvali-Gabadadze-Porrati) gravity \cite{DGP,luereview},
spacetime is five-dimensional, but energy-momentum is located on
a four-dimensional brane.  The action for gravity is
\begin{equation}
     S_{(5)} = -\frac{M^3}{16 \pi}\int d^5x \sqrt{-g} R -
      \frac{M_P^2}{16\pi} \int d^4x\sqrt{-g^{(4)}} R^{(4)},
\end{equation}
where $M$ is the five-dimensional Planck scale, $M_P$ the
observed four-dimensional Planck scale, $g$ and $R$ the bulk
metric and scalar curvature, and $g^{(4)}$ and $R^{(4)}$ those
on the brane.  On the brane, the gravitational potential due to
a point mass $m$ is $V\sim -G_{\rm brane} m/r$ at $r\ll r_0$, and
$V\sim -G_{\rm bulk}m /r^2$ at $r \gg r_0$, where $G_{\rm
bulk}=M^{-3}$ and $G_{\rm brane}=M_P^{-2}$ are five- and
four-dimensional Newton's constants, respectively, and
$r_0=M_P^2/2M^3$ is a cutoff scale that separates the ordinary
short-distance behavior from the new long-distance behavior.
Thus, gravity is weaker at large distances.  The theory admits
accelerating FRW solutions \cite{deffayet} that have $w_{\rm
eff}(z)=-1/(1+\Omega_m)$ and imply a crossover scale $r_0\sim
H_0^{-1}$.  Although it was originally believed that the model
would violate solar-system tests, in much the same way that
$1/R$ gravity does, the short-distance phenomenology of the
model is a bit more subtle \cite{luestarkman}.  The model leads
to a perihelion advance (in addition to the usual
general-relativistic one) for planetary orbits of $\Delta \phi
\sim 5(r^3/2r_0^2 r_g)^{1/2}$ with radius $r$, where $r_g=Gm$.
For values consistent with those required to explain cosmic
acceleration, the perihelion advance is consistent with
measurements, although, interestingly enough, possibly
detectable with future experiments.  As a classical theory of
gravity, DGP theory thus provides a theoretically sophisticated
arena for calculation and an interesting connection
between the cosmic acceleration and local tests of gravity.

Finally, it was suggested recently \cite{kolb} that cosmic acceleration
could be understood simply as a consequence of cosmological
inhomogeneities in general relativity, with{\it out} the
introduction of dark energy or alternative gravity.  This
proposal received a flurry of attention, but was then shown to
be unworkable \cite{chris}.

\subsection{Big Rip}

Prior to the advent of the data that indicated its existence,
hardly any theorist would have really believed in his/her
heart that there was a cosmological constant or some other sort
of negative-pressure dark energy.  The simplest phenomenological
models (i.e., the simplest single-field quintessence models), as
well as various energy conditions (an assortment of hypotheses
about the stress-energy properties allowed for matter), suggest
$w \geq -1$.  However, current data are consistent with
$w<-1$; for example, the latest WMAP data \cite{Spergel06}
indicate $w=-0.97^{+0.07}_{-0.09}$, centered near $w=-1$ but
consistent with $w<-1$.

\begin{table}
\caption{\label{tab:history}
The history and future of the Universe with $w=-3/2$ phantom energy.  }
\begin{tabular}{ll}
Time & Event\\
\hline  
$\sim 10^{-43}$~s  & Planck era \\ 
$\sim 10^{-36}$~s & Inflation \\
First Three Minutes & Light Elements Formed \\
$\sim 10^5$~yr & Atoms Formed \\
$\sim 1$~Gyr & First Galaxies Formed \\
$\sim 14$~Gyr & {\it Today} \\
$t_{\rm rip} - 1$~Gyr & Erase Galaxy Clusters\\
$t_{\rm rip} - 60$~Myr & Destroy Milky Way\\
$t_{\rm rip} - 3$~months & Unbind Solar System\\
$t_{\rm rip} - 30$~minutes & Earth Explodes\\
$t_{\rm rip} - 10^{-19}$~s & Dissociate Atoms\\
$t_{\rm rip} = 35$~Gyrs & Big Rip\\
\end{tabular}
\label{tab:table}
\end{table}

It is thus interesting to ask, what happens if dark energy is
phantom energy \cite{phantom}? i.e., what if it has an
equation-of-state parameter $w<-1$?  In this case, the
dark-energy density {\it increases} with time, and if $w$
remains less than $-1$, then it can be shown that the Universe
ends in a ``big rip,'' \cite{bigrip,mcinnes} a singularity in which the
Universe is stretched to infinite scale factor in finite time,
ripping everything in the Universe apart as it does so (see
Table \ref{tab:table}).  To
illustrate, let's imagine that the value of $w$ was $w=-1.5$.
In that case, the Universe,
currently about 14 billions year old, will stretch to infinite
size in about 20 billion years (with the constraints to $w$ from
WMAP, the onset of the big rip will occur later).  About a
billion years before that, galaxy clusters will be stripped
apart, and about 60 million years before, the Milky Way will
become dissociated.  Three months before the Big Rip, the Solar
System will be ripped apart, and then the Earth, about half an
hour before the end of time.  The final fraction of a second
will see atoms dissociated and ultimately, nuclei.

Although phantom energy is indeed somewhat fantastic, there have
been a number of exotic theoretical models for phantom energy,
based, e.g., on scalar-field models with higher-derivative
terms \cite{phantom,higherderivative}, or perhaps on supergravity
or higher-derivative gravity theories.  There have also been
models for $w<-1$ based on theories with higher dimensions
\cite{sahnishtanov}, strings \cite{frampton}, or the
AdS/CFT (for anti-de-Sitter space and conformal field theory)
correspondence \cite{mcinnes}.

\section{Conclusions}

Cosmology is in an exciting period.  What were until recently
wild theoretical speculations about the very earliest Universe
must now be considered very serious models.  Experiments that
were just until a few years ago ``futuristic'' have now been
completed, with spectacular success.  We have gone from being an
area in which the standard was order-of-magnitude estimates to a
precision science with elegant experiments with controlled
errors.  The results of the experiments have confirmed what was
long surmised---e.g., that most of the matter in the Universe is
nonbaryonic---and provided new surprises, such as the
accelerated expansion of the Universe.

In this brief review, I have discussed what we have learned from
CMB experiments, and then moved on to discuss the candidates we
have for dark matter and some of the ideas that have been
discussed for dark energy.  It must be realized that the CMB,
inflation, dark matter, and dark energy now occupy the attention
of a very significant fraction of the research enterprises of
both physics and astronomy.  There are thus an extraordinary
wealth of ideas as well as a plethora of detailed theoretical
calculations that I have not touched upon.  The interested
reader can use the reference list here as an introduction to
peruse the broader literature.

Where will cosmology go next?  We cannot say for sure.  One
obvious target is the CMB polarization due to inflationary
gravitational waves, which, as discussed above, may now---with
new CMB evidence for a scalar spectral index $n_s<1$---be likely
to be observable by next-generation experiments.  Then there are
dark-matter searches, which have been developing steadily in
sensitivity over the past few decades.  Again, a ``definitive''
experiment is hard to specify precisely, but experiments have
been steadily improving in sensitivity. It is conceivable that
within the next decade or two, we will probe most of the favored
supersymmetric parameter space.  Dark energy is here perhaps the
dark horse.  We are, theoretically, at a loss for really
attractive explanations for the dark energy.  The primary
observational question being addressed is whether it is a true
cosmological constant, or whether its density evolves with time.
However, this will be an experimental challenge.  And what
happens if it turns out to be consistent with a cosmological
constant?

\bigskip
{\bf Acknowledgments:} I thank Don York for a number of useful
suggestions.  This work was supported by DoE DE-FG03-92-ER40701,
NASA NNG05GF69G, and the Gordon and Betty Moore Foundation.

\section{Glossary of Technical Terms and
Acronyms\protect\footnote{Prepared in collaboration with Adrian Lee.}}

\noindent{\bf ACBAR (Arcminute Cosmology Bolometer Array Receiver).}  A bolometer-based
CMB temperature experiment that characterized the damping tail of CMB temperature
fluctuations.  It had a 16-element array and 4 arc-minute
resolution at 150 GHz\\
({\tt http://cosmology.berkeley.edu/group/swlh/acbar/}).

\noindent{\bf Acoustic peaks.}  Wiggles in the CMB
temperature and polarization power spectra that arise from acoustic
oscillations in the primordial baryon-photon fluid.

\noindent{\bf Adiabatic perturbations.}  Primordial density perturbations
in which the spatial distribution of matter is the same for all
particle species (photons, baryons, neutrinos, and dark
matter).  Such perturbations are produced by the simplest
inflation models.

\noindent{\bf AdS/CFT (Anti-de Sitter space/conformal field theory)
correspondence.}  A conjectured equivalence between string
theory in one space and a conformal gauge theory on the boundary
of that space.

\noindent{\bf AMANDA.} An astrophysical-neutrino observatory in deep
Antarctic ice\\
({\tt http://amanda.uci.edu}).

\noindent{\bf AMS (Alpha Magnetic Spectrometer).}  A NASA space-based
cosmic-ray-antimatter experiment ({\tt http://ams.cern.ch}).

\noindent{\bf APEX-SZ (Atacama Pathfinder EXperiment-Sunyaev-Zel'dovich).}  A
bolometer-based experiment designed to search for galaxy
clusters via the Sunyaev-Zel'dovich effect.  The 12-meter diameter APEX
telescope gives one arc-minute resolution at 150 GHz ({\tt http://bolo.berkeley.edu/apexsz/}).

\noindent{\bf Axion.}  A scalar particle that arises in the Peccei-Quinn
solution to the strong-CP problem.  If the axion has a mass near
$10^{-5}$ eV, then it could make up the dark matter.

\noindent{\bf Baksan experiment.}  A Russian underground
astrophysical-neutrino telescope ({\tt
http://www.inr.ac.ru/INR/Baksan.html}).

\noindent{\bf BICEP (Background Imaging of Cosmic Extragalactic Polarization).}  
A  bolometer-based CMB polarization experiment sited at the South Pole.  It uses a 
small refractive telescope to achieve 0.6 degree resolution at
150 GHz\\
({\tt http://www.astro.caltech.edu/$\sim$lgg/bicep\_front.htm}).

\noindent{\bf Baryons.}  In cosmology, this term refers to ordinary
matter composed of neutrons, protons, and electrons.

\noindent{\bf BBN (Big-bang nucleosynthesis).}  The theory of the assembly of
light nuclei from protons and neutrons a few seconds to minutes
after the big bang.

\noindent{\bf BBO (Big Bang Observer).}  A mission concept, currently under
study, for a post-LISA space-based gravitational-wave
observatory designed primarily to seek inflationary
gravitational waves\\
({\tt http://universe.nasa.gov/program/bbo.html}).

\noindent{\bf BESS (Balloon-borne Experiment with a Superconducting
Spectrometer).}  A Japanese-US collaborative series of
balloon-borne experiments to measure antimatter in cosmic rays\\
({\tt http://www.universe.nasa.gov/astroparticles/programs/bess/}).

\noindent{\bf Big rip.}  A possible end fate for the Universe in which
the Universe expands to infinite size in finite time, ripping
everything apart as it does so.

\noindent{\bf Boltzmann equations.}  Equations for the evolution of the
momentum distributions for various particle species (e.g.,
baryons, photons, neutrinos, and dark-matter particles).

\noindent{\bf BOOMERanG.}  A balloon-borne CMB-fluctuation experiment
that reported in 2000 the first measurement of acoustic-peak
structure in the CMB.  It used a bolometer array and had 10 arc-minute
resolution at 150 GHz\\
({\tt http://cmb.phys.cwru.edu/boomerang}).

\noindent{\bf Brane} or {\bf $p$-brane}.  A $p$-dimensional subspace of
some higher-dimensional subspace.  As an example, in some string
theories, there may be many extra dimensions, but standard-model
fields are restricted to lie in a 4-dimensional volume that is
our $3+1$-dimensional spacetime.

\noindent{\bf CACTUS.} A heliostat array for $>40$ GeV gamma-ray
astronomy\\
 ({\tt http://ucdcms.ucdavis.edu/solar2}).

\noindent{\bf CAPRICE (Cosmic AntiParticle Ring Imaging Cherenkov
Experiment).}  A 1994 balloon-borne cosmic-ray-antimatter
experiment\\
 ({\tt http://www.roma2.infn.it/research/comm2/caprice}).
 
\noindent{\bf CBI (Cosmic Background Imager).}  An interferometric CMB
telescope designed to measure the smallest-angular-scale
structure of the CMB\\
({\tt http://www.astro.caltech.edu/$\sim$tjp/CBI}).

\noindent{\bf CAPMAP (Cosmic Anisotropy Polarization MAPper).}  A CMB polarization
experiment using the Lucent Technologies 7-meter diameter
 telescope at Crawford Hill NJ and coherent detectors\\
({\tt http://quiet.uchicago.edu/capmap/}).

\noindent{\bf CDMS (Cryogenic Dark Matter Search).}  A U.S. experiment
designed to look for WIMPs ({\tt http://cdms.berkeley.edu}).

\noindent{\bf CELESTE.} A heliostat array for $\sim100$ GeV gamma-ray
astronomy.

\noindent{\bf CMB (Cosmic microwave background).}  A 2.7 K gas of
thermal radiation that permeates the Universe, a relic of the
big bang.

\noindent{\bf CMBPOL.}  A mission concept, currently under study, for a
post-Planck CMB satellite experiment designed primarily to
search for inflationary gravitational waves.

\noindent{\bf COBE (Cosmic Background Explorer).} A NASA satellite flown
from 1990--1993 with several experiments designed to measure
the properties of the CMB.  John Mather and George Smoot, two of
the leaders of COBE, were awarded the 2006 Nobel prize for
physics for COBE\\
 ({\tt http://lambda.gsfc.nasa.gov/product/cobe}).

\noindent{\bf Cosmic jerk.}  A parameter that quantifies the time
variation of the cosmic acceleration.

\noindent{\bf Cosmic shear (CS).}  Gravitational lensing of distant
cosmological sources by cosmological density perturbations along
the line to those sources.

\noindent{\bf Cosmological constant ($\Lambda$).}  An extra term in the
Einstein equation that quantifies the gravitating mass density
of the vacuum.

\noindent{\bf Critical density.}  The cosmological density required for a
flat Universe.  If the density is higher than the
critical density, then the Universe is closed, and if it is
smaller, then it is open.

\noindent{\bf DAMA.}  An Italian experiment designed to look for
WIMPs\\
({\tt http://people.roma2.infn.it/$\sim$dama/web/home.html}).

\noindent{\bf Dark energy (DE).}  A form of negative-pressure matter that
fills the entire Universe.  It is postulated to account for the
accelerated cosmological expansion.

\noindent{\bf Dark matter (DM).}  The nonluminous matter required to account
for the dynamics of galaxies and clusters of galaxies.  The
preponderance of the evidence suggests that dark matter is not
made of baryons, and it thus often referred to as ``nonbaryonic
dark matter.''  The nature of dark matter remains a mystery.

\noindent{\bf DASI (Degree Angular Scale Interferometer).}  An interferometric 
CMB experiment sited at the South Pole that characterized the acoustic peaks in the CMB power
spectrum and first detected the E-mode polarization in the CMB\\
({\tt http://astro.uchicago.edu/dasi/}).

\noindent{\bf DECIGO (Deci-hertz Interferometer Gravitational Wave
Observatory).}  A mission concept, currently under study in
Japan, for an even more ambitious version of BBO.

\noindent{\bf DGP (Dvali-Gabadadze-Porrati) gravity.}  A theory for
gravity, that may explain cosmic acceleration, based on the
introduction of one extra spatial dimension.

\noindent{\bf Dirac neutrino.}  A type of neutrino that has an
antiparticle.

\noindent{\bf DMR (Differential Microwave Radiometer).}  An experiment
on COBE that measured temperature fluctuations in the CMB\\
({\tt http://lambda.gsfc.nasa.gov/product/cobe}).

\noindent{\bf EDELWEISS.}  A French experiment designed to look for
WIMPs\\
({\tt http://edelweiss.in2p3.fr}).

\noindent{\bf EGRET (Energetic Gamma Ray Experiment Telescope).}   A
high-energy gamma-ray experiment flown aboard NASA's Compton
Gamma-Ray Observatory in the early 1990s\\
({\tt http://cossc.gsfc.nasa.gov/docs/cgro/cossc/EGRET.html}).

\noindent{\bf Einstein's equations.}  The equations of general
relativity.

\noindent{\bf Electroweak (EW) phase transition.}  The phase transition at a
temperature $\sim100$ GeV that breaks the electroweak symmetry
at low energies to distinct electromagnetic and weak
interactions.

\noindent{\bf Friedmann equation.} The general-relativistic equation that
relates the cosmic expansion rate to the cosmological energy
density.

\noindent{\bf Friedman-Robertson-Walker (FRW) spacetime.}  The spacetime that
describes a homogeneous isotropic Universe.

\noindent{\bf Galaxy clusters.}  Gravitationally bound systems of
hundreds to thousands of galaxies.

\noindent{\bf General relativity (GR).} Einstein's theory that combines
gravity with relativity.

\noindent{\bf GLAST (Gamma Ray Large Area Space Telescope).}  A NASA
telescope, to be launched within a year, for high-energy
gamma-ray astronomy\\
({\tt http://www-glast.stanford.edu}).

\noindent{\bf Grand-unified theories (GUTs).}  Gauge theories that unify
that electroweak and strong interactions at an energy
$\sim10^{16}$ GeV.

\noindent{\bf Gravitational lensing.}  The general-relativistic bending
of light by mass concentrations.

\noindent{\bf Gravitational waves (GWs).}  Propagating disturbances, which
arise in general relativity, in the gravitational field,
analogous to electromagnetic waves (which are propagating
disturbances in the electromagnetic field).

\noindent{\bf Hawking radiation.}  Radiation emitted, as a result of
quantum-mechanical processes, from a black hole.

\noindent{\bf HEAT (High Energy Antimatter Telescope).}  A balloon-borne
cosmic-ray-antimatter telescope from the 1990s.

\noindent{\bf HESS (High Energy Stereoscopic System).}  A ground-based
air Cerenkov telescope for GeV--TeV gamma-ray astronomy\\
({\tt http://www.mpi-hd.mpg.de/hfm/HESS/HESS.html}).

\noindent{\bf Hubble constant.}  The constant of proportionality between
the recessional velocity of galaxies and their distance.  The
Hubble constant is also the expansion rate.  When used in this
context, the term is a misnomer, as the expansion rate varies
with time.

\noindent{\bf IceCube.} An astrophysical-neutrino observatory (a
successor to AMANDA) now being built at the South Pole ({\tt
http://icecube.wisc.edu}).

\noindent{\bf IMAX (Isotopie Matter Antimatter Telescope).}  A 1992
balloon-borne cosmic-ray-antimatter telescope ({\tt
http://www.srl.caltech.edu/imax.html}).

\noindent{\bf Inflation.}  A period of accelerated expansion in the early
Universe postulated to account for the isotropy and homogeneity
of the Universe.

\noindent{\bf Inflationary gravitational waves (IGWs).}  A cosmological
background of gravitational waves produced via quantum processes
during inflation.

\noindent{\bf IMB (Irvine-Michigan-Brookhaven) experiment.}
A U.S.
underground detector designed originally to look for proton
decay, but used ultimately (from 1979--1989) as an
astrophysical-neutrino detector\\
({\tt http://www-personal.umich.edu/$\sim$jcv/imb/imb.html}).

\noindent{\bf JDEM (Joint Dark Energy Mission).}  A space mission in
NASA's roadmap that aims to study the cosmic acceleration\\
({\tt http://universe.nasa.gov/program/probes/jdem.html}).

\noindent{\bf Kaluza-Klein (KK) modes.}  Excitations of a fundamental
field in extra dimensions in a theory with extra dimensions.
These modes appear as massive particles in our 3+1-dimensional
spacetime.

\noindent{\bf Kamiokande and Super-Kamiokande.}  A Japanese underground
astrophysical-neutrino telescope (and proton-decay experiment)
and its successor\\
({\tt http://www-sk.icrr.u-tokyo.ac.jp/sk/index.html}).

\noindent{\bf Large extra dimensions.}  A currently popular idea in
particle theory that the Universe may contain more spatial
dimensions than the three that we see, and that the additional
dimensions may be large enough to have observable consequences.

\noindent{\bf Large-scale structure (LSS).}  The spatial distribution of
galaxies and clusters of galaxies in the Universe.

\noindent{\bf Laser Interferometric Space Antenna (LISA).}  A satellite
experiment planned by NASA and ESA to detect gravitational waves
from astrophysical sources\\
({\tt http://lisa.nasa.gov}).

\noindent{\bf LEP (Large Electron-Positron) Collider.}  The
electron-positron collider at CERN (European Center for Nuclear
Research) which from 1989 to 2000 tested with exquisite
precision the Standard Model.

\noindent{\bf LHC (Large Hadron Collider).}  The successor the LEP at
CERN, the LHC will be (starting November 2007) a proton-proton
collider, and the world's most powerful particle accelerator.

\noindent{\bf LSP (Lightest superpartner).}  The lightest supersymmetric
particle (and a candidate WIMP) in supersymmetric extensions of
the Standard Model.

\noindent{\bf LIGO (Laser Interferometric Gravitational-Wave
Observatory).}  An NSF experiment, currently operating, designed
to detect gravitational waves from astrophysical sources ({\tt
http://www.ligo.caltech.edu}).

\noindent{\bf Local Group.}  The group of galaxies that the Milky Way
belongs to.

\noindent{\bf LSST (Large Synoptic Survey Telescope).}  A proposed
wide-field survey telescope ({\tt
http://www.lsst.org/lsst\_home.shtml}).

\noindent{\bf Lyman-alpha forest} or {\bf Ly-$\alpha$ forest}.  The
series of absorption features, in the spectra of
distant quasars, due to clouds of neutral hydrogen along the
line of sight.

\noindent{\bf Majorana neutrino.}  A type of neutrino that is its own
antiparticle.

\noindent{\bf MACRO (Monopoles and Cosmic Ray Observatory).} An
underground astrophysical-neutrino telescope
(and proton-decay experiment) that ran at the Gran Sasso
Laboratory in Italy from 1988 to 2000.

\noindent{\bf MASS (Matter Antimatter Superconducting Spectrometer).}  A
1989--1991 balloon-borne cosmic-ray-antimatter telescope\\
({\tt http://people.roma2.infn.it/$\sim$aldo//mass.html}).

\noindent{\bf MAT/TOCO (Mobile Anisotropy Telescope on Cerro TOCO).}  A CMB experiment 
using coherent detectors that gave early results on the location 
of the first acoustic peak in the CMB angular power spectrum  \\
({\tt http://www.physics.princeton.edu/cosmology/mat/}).

\noindent{\bf MAXIMA (Millimeter Anisotropy eXperiment Imaging Array).}  
A balloon-borne experiment that reported in 2000
measurements of temperature fluctuations on degree angular
scales.  It had a 16 element bolometer array operated at 100 mK and 10 arc-minute beams at
150 GHz \\
({\tt http://cosmology.berkeley.edu/group/cmb}).

\noindent{\bf MAXIPOL.}  A balloon-borne CMB polarization experiment based
on the MAXIMA experiment  \\
({\tt http://groups.physics.umn.edu/cosmology/maxipol/}).

\noindent{\bf Naturalness problem.}  In grand-unified theories without
supersymmetry, the parameter that controls the EW
symmetry-breaking scale must be tuned to be extremely small.

\noindent{\bf NET (Noise-equivalent temperature).}  A quantity that
describes the sensitivity (in units of $\mu$K$\sim$$\sqrt{\rm sec}$)
of a detector in a CMB experiment.

\noindent{\bf Neutralino.}  The superpartner of the photon and $Z^0$ and
Higgs bosons, and an excellent WIMP candidate in supersymmetric
extensions of the standard model.

\noindent{\bf PAMELA.}  A space-based cosmic-ray-antimatter experiment
flown in 2006\\
({\tt http://wizard.roma2.infn.it/pamela}).

\noindent{\bf Peccei-Quinn mechanism.}  A mechanism, involving the
introduction of a new scalar field, that solves the strong-CP
problem.

\noindent{\bf Phantom energy.} An exotic form of dark energy that is
characterized by an equation-of-state parameter $w<-1$.

\noindent{\bf Planck satellite.}  A collaborative NASA/ESA satellite
experiment aimed to measure temperature fluctuations in the CMB
with even more precision and sensitivity than WMAP\\
({\tt http://www.rssd.esa.int/Planck}).

\noindent{\bf Planck-scale physics.}  A colloquial term that refers to
quantum gravity or string theory.

\noindent{\bf POLARBeaR (POLARization of the Background Radiation).}  
A planned bolometer-based CMB polarization experiment to be sited in Chile\\
({\tt http://bolo.berkeley.edu/polarbear/index.html}).

\noindent{\bf Primordial density perturbations} or sometimes just {\bf
primordial perturbations.}  The small-amplitude primordial density
inhomogeneities (which may have arisen during inflation) that
were amplified via gravitational instability into the large-scale
structure we see today.

\noindent{\bf Pseudo-Nambu-Goldstone boson.}  A nearly massless scalar
particle that arises in a theory with an explicitly broken
global symmetry.

\noindent{\bf PVLAS.}  A laser experiment designed to look for the vacuum
magnetic birefringence predicted in quantum electrodynamics\\
({\tt http://www.ts.infn.it/physics/experiments/pvlas/pvlas.html}).

\noindent{\bf Q-balls.}  Extended objects, composed of a
a spinning scalar field, that appear in scalar field theories
with a $U(1)$ symmetry (i.e., a cylindrical symmetry in the
internal space).

\noindent{\bf QCD (Quantum chromodynamics).}  The theory of the strong
interactions that confine quarks inside protons and neutrons.

\noindent{\bf QuaD (Q and U Extra-galactic Sub-Millimetre Telescope and DASI).}  
A bolometer-based CMB polarization experiment at the South Pole.  It has 4 arc-minute
resolution at 150 GHz\\
({\tt http://www.stanford.edu/$\sim$schurch/quad.html}).

\noindent{\bf Quantum gravity.}  A term that refers to a theory---still
to be determined but widely believed to be string theory---that
unifies quantum mechanics and gravity.

\noindent{\bf Quark-hadron phase transition} or {\bf QCD phase
transition}.  The transition at temperature $\sim100$ MeV at
which quarks are first bound into protons and neutrons.

\noindent{\bf Quintessence.} A mechanism postulated to explain cosmic
acceleration by the displacement of a scalar field (the
quintessence field) from the minimum of its potential.

\noindent{\bf Recombination.}  The formation of atomic hydrogen and
helium at a redshift $z\simeq1100$.

\noindent{\bf Redshift ($z$).}  The recessional velocity of a galaxy divided by
the speed of light.  The redshift is used as a proxy for
distance or time after the big bang, with higher redshift
indicating larger distances and earlier times.

\noindent{\bf SKA (Square-Kilometer Array).}  A large radio-telescope
array planned by NSF ({\tt http://www.skatelescope.org}).

\noindent{\bf SNAP (Supernova Acceleration Probe).}  A proposed
space-based telescope dedicated to measuring the cosmic
expansion history ({\tt http://snap.lbl.gov}).

\noindent{\bf SPIDER.} A balloon-borne bolometer-based CMB polarization
experiment with six refractive telescopes\\
({\tt http://www.astro.caltech.edu/$\sim$lgg/spider\_front.htm
}).

\noindent{\bf Spintessence.} A variant of quintessence in which the
scalar field is taken to be complex with a $U(1)$ symmetry.

\noindent{\bf SPUD (Small Polarimeter Upgrade for Dasi ).}  A proposed
CMB experiment to be attached to the DASI mount at the South Pole.

\noindent{\bf STACEE (Solar Tower Atmospheric Cerenkov Effect
Experiment).}   A ground-based air Cerenkov telescope designed
to detect gamma rays in the $\sim100$ GeV range ({\tt
http://www.astro.ucla.edu/$\sim$stacee}).

\noindent{\bf Standard Model (SM).}  The theory of strong, weak, and
electromagnetic interactions.

\noindent{\bf String theory.}  A theory that postulates that all
elementary particles are excitations of fundamental strings.
The aim of such theories is to unify the strong and
electroweak interactions with gravity at the {\bf Plank scale},
an energy scale $\sim10^{19}$ GeV.

\noindent{\bf Strong-CP problem.}  Although the strong interactions are
observed to be parity conserving, there is nothing in QCD that
demands that parity be conserved.

\noindent{\bf Supersymmetry (SUSY).}  A symmetry between fermions and bosons
postulated primarily to solve the naturalness problem.  It is an
essential ingredient in many theories for new physics beyond the
Standard Model.

\noindent{\bf Triangle anomaly.}  A coupling, mediated by the exchange of
virtual fermions, between a scalar particle and two photons.
This coupling is responsible for neutral-pion decay to two
photons.

\noindent{\bf TS93.} A 1993 balloon-borne cosmic-ray-antimatter
telescope\\
({\tt http://people.roma2.infn.it/$\sim$aldo//ts93.html}).

\noindent{\bf Universal extra dimensions (UED).}  A class of theories for
new physics at the electroweak scale in which the Universe has
extra large dimensions in which standard-model fields propagate.

\noindent{\bf Vacuum energy.}  The energy of free space.

\noindent{\bf VERITAS (Very Energetic Radiation Imaging Telescope Arrays
System).}  A ground-based air Cerenkov telescope for GeV--TeV
gamma-ray astronomy ({\tt http://veritas.sao.arizona.edu}).

\noindent{\bf VSA (Very Small Array).}  A ground-based CMB interferometer that
is sited in the Canary Islands.  It is sensitive to a wide range
 of angular scales with a best resolution of 10 arc-minute\\
({\tt http://www.mrao.cam.ac.uk/telescopes/vsa/index.html}).

\noindent{\bf WIMP (Weakly-interacting massive particle).}  A dark-matter
candidate particle that has electroweak interactions with
ordinary matter.  Examples include massive neutrinos,
supersymmetric particles, or particles in models with universal
extra dimensions.

\noindent{\bf WMAP (Wilkinson Microwave Anisotropy Probe).}  A NASA
satellite launched in 2001 to measure, with better sensitivity
and angular resolution than DMR, the temperature
fluctuations in the CMB\\
({\tt http://map.gsfc.nasa.gov}).

\noindent{\bf ZEPLIN} An experiment designed to look for WIMPs.

\nonumsection{References}

\end{document}